\documentclass[journal]{IEEEtran}
\usepackage[utf8]{inputenc}
\usepackage[english]{babel}
\usepackage{makeidx}
\usepackage{graphicx}
\usepackage[hidelinks]{hyperref}
\usepackage{amsmath}
\usepackage[inline]{enumitem}
\usepackage{multirow}
\usepackage{subcaption}
\usepackage{caption}
\usepackage{color}
\usepackage{booktabs}
\usepackage{siunitx}
\usepackage{cite}
\usepackage{algpseudocode}
\usepackage{algorithmicx}
\usepackage{array}
\usepackage{float}
\usepackage{todonotes}
\usepackage[keeplastbox]{flushend}
\usepackage[linesnumbered,ruled,vlined]{algorithm2e}
\usepackage{amsfonts}
\usepackage{threeparttable}
\usepackage{ragged2e}
\DeclareUnicodeCharacter{2061}{}

\algnewcommand\algorithmicforeach{\textbf{for each}}
\algdef{S}[FOR]{ForEach}[1]{\algorithmicforeach\ #1\ \algorithmicdo}

\begin{document}
\title{Clinically Verified Hybrid Deep Learning System for Retinal Ganglion Cells Aware Grading of Glaucomatous Progression}

\author{Hina Raja\textsuperscript{*}, 
        Taimur~Hassan\textsuperscript{*}\textsuperscript{$\dagger$}, 
        Muhammad~Usman~Akram,~\IEEEmembership{Member,~IEEE,}
        Naoufel~Werghi,~\IEEEmembership{Senior~Member,~IEEE}% <-this % stops a space
\thanks{\noindent Copyright $\copyright$ 2020 IEEE. Personal use of this material is permitted. However, permission to use this material for any other purposes must be obtained from the IEEE by sending an email to pubs-permissions@ieee.org. }
\thanks{\noindent This work is supported by a research fund from Khalifa University: Ref: CIRA-2019-047.}% <-this % stops a space
\thanks{\noindent H. Raja and M. U. Akram are with the Department of Computer and Software Engineering, National University of Sciences and Technology, Islamabad, Pakistan.}% <-this % stops a space
\thanks{\noindent T. Hassan and N. Werghi are with the Center for Cyber-Physical Systems (C2PS), Department of Electrical Engineering and Computer Sciences, Khalifa University, Abu Dhabi, United Arab Emirates.}
\thanks{\noindent \textsuperscript{*}Co-first authors \textsuperscript{$\dagger$}Corresponding author, Email: taimur.hassan@ku.ac.ae}}

\markboth{IEEE Transactions on Biomedical Engineering, October 2020}
{Raja \MakeLowercase{\textit{et al.}}: Clinically Verified Hybrid Deep Learning System for Retinal Ganglion Cells Aware Grading of Glaucomatous Progression}
%

\maketitle

\begin{abstract}
\textit{Objective:} Glaucoma is the second leading cause of blindness worldwide. Glaucomatous progression can be easily monitored by analyzing the degeneration of retinal ganglion cells (RGCs). Many researchers have screened glaucoma by measuring cup-to-disc ratios from fundus and optical coherence tomography scans. However, this paper presents a novel strategy that pays attention to the RGC atrophy for screening glaucomatous pathologies and grading their severity. 
\textit{Methods:} The proposed framework encompasses a hybrid convolutional network that extracts the retinal nerve fiber layer, ganglion cell with the inner plexiform layer and ganglion cell complex regions, allowing thus a quantitative screening of glaucomatous subjects. Furthermore, the severity of glaucoma in screened cases is objectively graded by analyzing the thickness of these regions. 
\textit{Results:} The proposed framework is rigorously tested on publicly available Armed Forces Institute of Ophthalmology (AFIO) dataset, where it achieved the $F_1$ score of 0.9577 for diagnosing glaucoma, a mean dice coefficient score of 0.8697 for extracting the RGC regions and an accuracy of 0.9117 for grading glaucomatous progression. Furthermore, the performance of the proposed framework is clinically verified with the markings of four expert ophthalmologists, achieving a statistically significant Pearson correlation coefficient of 0.9236.
\textit{Conclusion:} An automated assessment of RGC degeneration yields better glaucomatous screening and grading as compared to the state-of-the-art solutions. 
\textit{Significance:} An RGC-aware system not only screens glaucoma but can also grade its severity and here we present an end-to-end solution that is thoroughly evaluated on a standardized dataset and is clinically validated for analyzing glaucomatous pathologies.

\end{abstract}


\IEEEpeerreviewmaketitle

\section{Introduction}

\begin{figure}[t]
    \includegraphics[scale=0.1755]{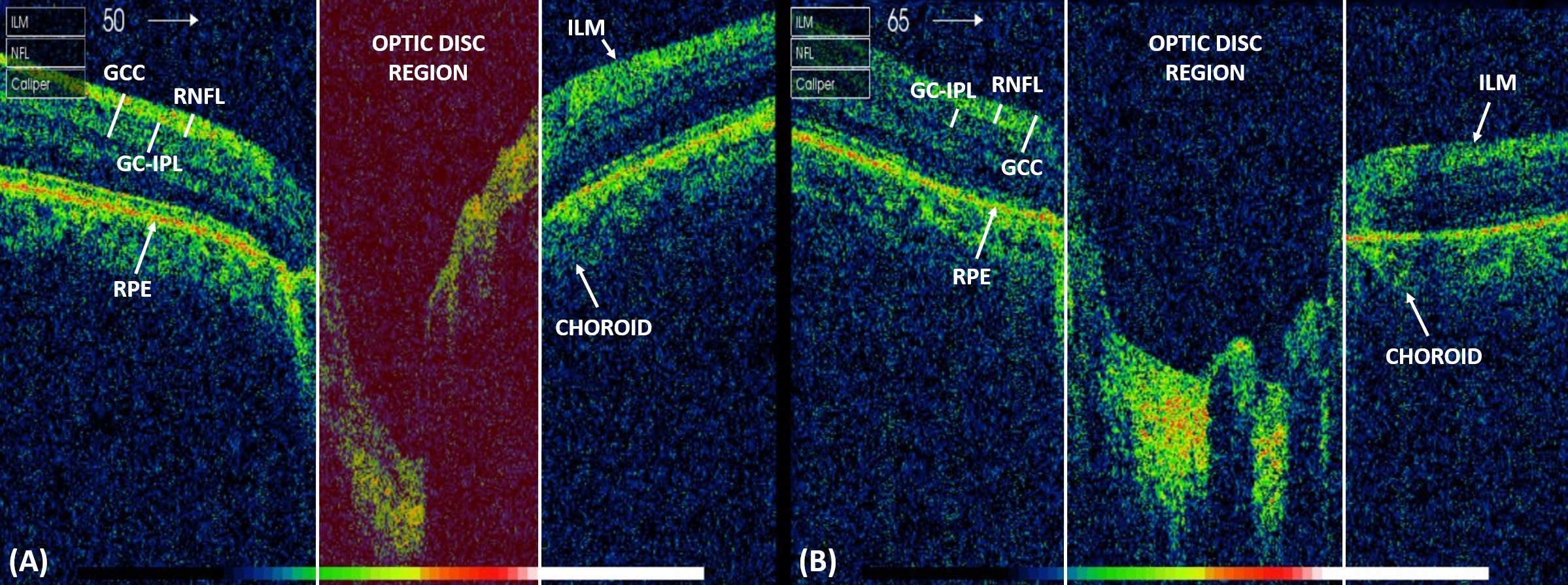}
\caption{\small Optic nerve head (ONH) OCT scan depicting (A) healthy and (B) glaucomatous pathology. Inner Limiting Membrane (ILM), RNFL, GC-IPL, GCC, and Retinal Pigment Epithelium (RPE) are highlighted along with the choroidal and optic disc region. The thinning of RNFL, GC-IPL, and the GCC regions can be observed in the glaucomatous scan (B) as compared to the healthy one (A). Both scans are taken from publicly available Armed Forces Institute of Ophthalmology (AFIO) dataset \cite{Raja2020DIB}. }
\centering
\label{fig:fig1}
\end{figure}

\IEEEPARstart{T}{he} human retina is composed of various nerve cells. Among these, retinal ganglion cells (RGCs) are responsible for transmitting visual information to the brain. The axons of these ganglion cells collectively form the retinal never fiber layer (RNFL), their cell bodies are enclosed in the ganglion cell layer (GCL) and the inner plexiform layer (IPL) embodies their dendrites. The composition of these layers is commonly termed as ganglion cell complex (GCC). Glaucoma (a progressive optic neuropathy) severely degrades these RGCs, resulting in a thinning of RNFL, ganglion cell with the inner plexiform layer (GC-IPL), and the GCC profiles, as shown in Figure \ref{fig:fig1}. This RGC atrophy can cause permanent visual impairments and even blindness if left untreated  \cite{weinreb2014JAMA}. 
Clinically, glaucoma can be identified through fundus and optical coherence tomography (OCT) based examinations. OCT imaging is generally preferred by the clinicians over other modalities due to its objective assessments of early and advanced staged glaucoma where early glaucoma refers to the condition when the RGCs start to degenerate due to increased intraocular pressure \cite{Burgoyne2005ONH}. However, the progression of RGC dysfunction leads towards the advanced glaucomatous stage where the total cupping of the optic nerve and severe visual impairments can be observed.  

\section{Related Work}\label{sec:RelatedWork}
\noindent Many researchers have worked on diagnosing glaucoma from retinal OCT images. These studies either emphasize the clinical significance of retinal OCT examination for analyzing glaucomatous severity, or they propose OCT-based autonomous systems for analyzing the glaucomatous pathologies. 
\subsection{Clinical Studies}
\noindent Development in retinal imaging modalities (especially OCT) is making rapid strides in providing the objective visualization of early, mild, and severe ocular complications \cite{Hassan2015IST}, particularly for the glaucoma \cite{majoor2019TVST}, the second leading cause of irreversible blindness worldwide. 
Moreover, the detection and monitoring of glaucoma by measuring the velocity of RNFL thickness loss has been significantly highlighted in many recent state-of-the-art studies \cite{grewal2012Glaucoma, Gracitelli2015Ophthalmol}.  Leung et al. \cite{leung2012Ophthalmology} demonstrated the importance of RNFL thickness (generated through OCT and visual field tests) in determining the retinal pathological variations within different glaucomatous stages.
Ojima et al. \cite{ojima2007Ophthalmol} signified the importance of RNFL thickness and macular volume, and declared that RNFL thickness has higher diagnostic power than a complete macular volume to detect glaucoma. 
Furthermore, Medeiros et al. \cite{Medeiros2012IOVS} evaluated RGC loss using standard automated perimetry (SAP) and spectral-domain OCT (SD-OCT) examinations. 
They observed that the early pathological degeneration of RGC results in the drastic thinning of RNFL as compared to the RGC changes in the late glaucomatous stages. 
Likewise, El-Naby et al. \cite{El-Naby2014Egyptian} extracted the RNFL thickness from SD-OCT scans and compared it with the VF sensitivity to observe their correlation in screening primary open-angle glaucoma. They concluded that the mean RNFL thickness obtained through SD-OCT imagery is a very good indicator of screening glaucomatous subjects and also for monitoring the progression and severity of the disease.

\subsection{Automated Glaucomatous Analysis}
\noindent Initial methods developed for glaucomatous screening analyze cup-to-disc ratios from macula-centered and disc-centered fundus images \cite{Khalil2020Wiley, Sun2018Localizing, Chen2015Glaucoma, Cheng2013Superpixel, Fu2018DiscAware}. However, observing the degeneration of RGCs through optic nerve head (ONH) SD-OCT scans can provide a superior and objective indication of early glaucoma, resulting in the timely prevention of non-recoverable visual impairments. Furthermore, due to the unprecedented clinical significance of retinal OCT examination \cite{majoor2019TVST, grewal2012Glaucoma, Gracitelli2015Ophthalmol}, many researchers have developed autonomous systems to objectively screen glaucoma (especially in early-stage) using retinal OCT scans \cite{Khalil2018Access}.
Moreover, Almobarak et al. \cite{Almobarak2014IOVS} manually segmented ONH structures from SD-OCT scans to analyze pathological variations in healthy and glaucomatous pathologies. Kromer et al. \cite{Kromer2017Ophthalmology} extracted eight retinal boundaries from 40 SD-OCT scans of healthy subjects using curve regularization. In \cite{Duan2018Access}, a generated model was presented to segment retinal layers from OCT images using a group-wise curve alignment. Niu et al. \cite{Niua2014CBM} proposed an automated method to segment the six retinal layers using correlation smoothness constraint and dual gradients. Apart from this, several methods have been proposed to quantify retinal layer thickness from SD-OCT scans depicting normal  \cite{Kafieh2015Hindawi, Bagci2007LSSAW}, and abnormal retinal pathologies  \cite{Abdellatif2019Hindawi, Hassan2016AO, Hassan2019CBM, Hassan2018Access, Chiu2015BOE}. Ometto et al. \cite{Ometto2019Trans} presented ReLayer, an automated framework to segment and estimate the thickness of ILM, inner/outer segment, and RPE from OCT scans to monitor retinal abnormalities. 
Gao et al. \cite{Gao2015PLOSONE} extracted retinal layers through graph decomposition from Topcon SD-OCT scans and evaluated the mean macular thickness of RNFL regions. Afterward, they compared their obtained results with the thickness measurements from Topcon’s built-in layer extraction framework. Likewise, Mayer et al. \cite{Mayer2010BOE} proposed an automated framework for extracting the retinal layers and computing the RNFL thickness by minimizing the energy obtained through scan gradients, local and regional smoothing filters. They validated their framework on a dataset containing both normal and glaucomatous affected OCT scans, and achieved a mean RNFL thickness of 94.1$\pm$11.7$\mu m$ and 65.3$\pm$15.7$\mu m$, respectively for normal and glaucomatous pathologies. In addition to this, many researchers have proposed computer-aided diagnostic systems to diagnose glaucomatous pathologies from fundus \cite{Khalil2017IET, Sun2018Localizing, Chen2015Glaucoma, Cheng2013Superpixel}, OCT \cite{Khalil2018Access} and fused fundus and OCT imagery \cite{Khalil2020Wiley}.
More recently, deep learning has been applied to analyze the glaucomatous pathologies through segmentation-free retinal layers extraction framework \cite{Mariottoni2020ScientificReports}. Zang et al. \cite{Zang2019BOE} used a convolutional neural network (CNN) and graph search to delineate the retinal boundaries and the optic disc region from ONH SD-OCT scans of normal and glaucomatous subjects, achieving the overall dice coefficient of 0.91$\pm$0.04. Maetschke et al. \cite{Maetschke2019PLOSONE} highlighted the significance of RNFL, GC-IPL profiles for diagnosing, and monitoring glaucoma progression and used the 3D CNN model to classify healthy and glaucomatous ONH SD-OCT scans. They outperformed conventional machine learning approaches by achieving the area under the curve ($AUC$) score of 0.94. Furthermore, Devalla et al. \cite{Sripad2018BOE} proposed a dilated-residual U-Net architecture (DRUNET) for the extraction of six ONH tissues from SD-OCT scans to aid experts in analyzing glaucomatous progression. DRUNET achieved the overall dice coefficient of 0.91$\pm$0.05 when accessed against manual tissue segmentation done by the expert observer. In addition to this, a joint retinal segmentation and classification pipeline was proposed in \cite{Wang2019BOE} to analyze healthy and glaucomatous pathologies from 1,004 locally acquired circular OCT scans, and also a severe diabetic macular edema (DME) pathology from selective 110 macular OCT scans of Duke dataset \cite{Chiu2015BOE}.  

\begin{figure*}[htb]
    \includegraphics[width=1\linewidth]{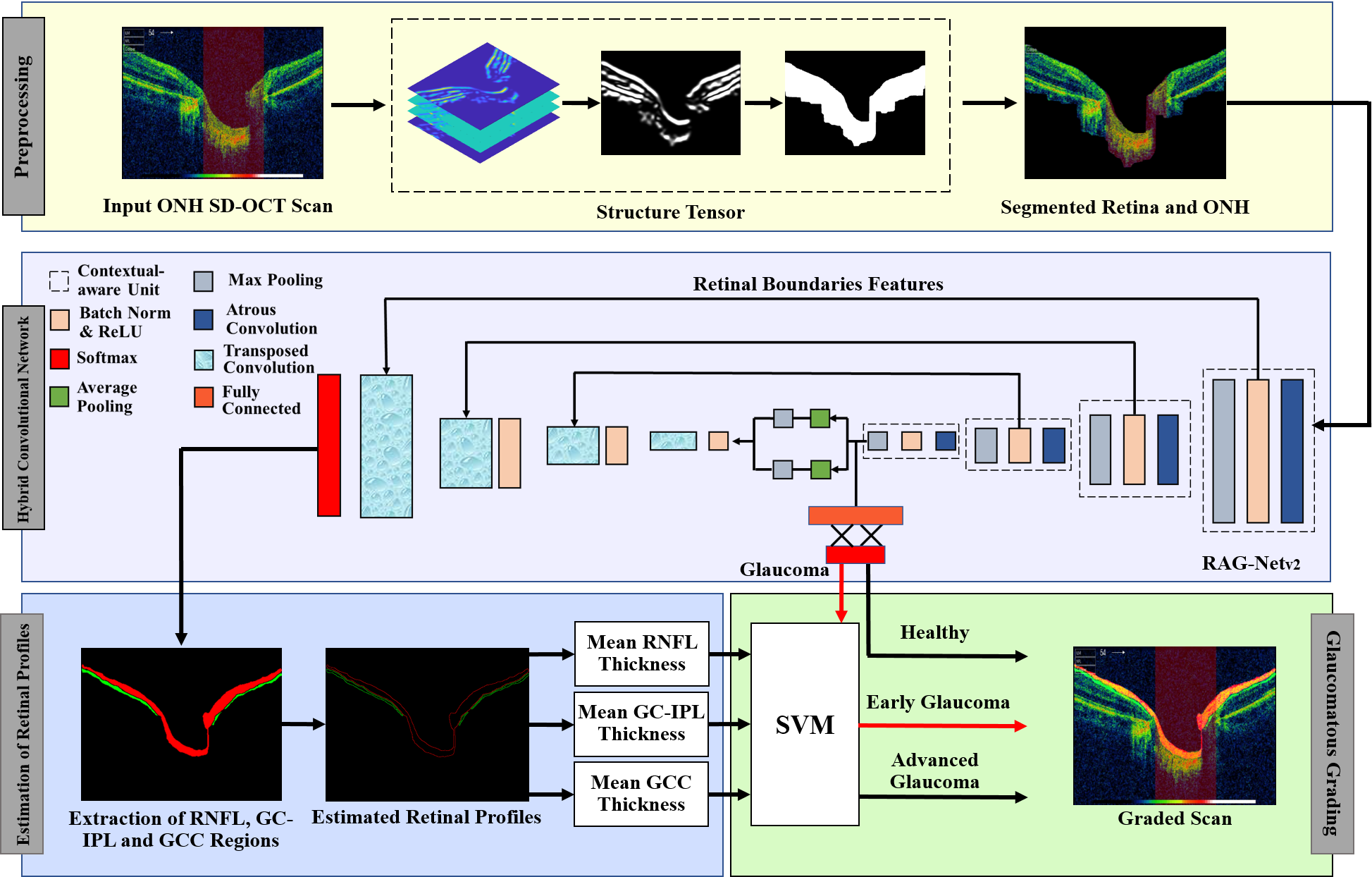}
\caption{\small The block diagram of the proposed framework. First of all, the input scan is preprocessed to remove the background noise and vendor annotations. Afterward, the processed scan is passed to the hybrid convolutional network (RAG-Net\textsubscript{v2}) for the simultaneous extraction of RNFL, GC-IPL, and GCC regions, and its classification against glaucoma. The screened glaucomatous scan is further graded by SVM based on the RGC atrophy observed through RNFL, GC-IPL, and GCC thickness profiles. }
\centering
\label{fig:fig2}
\end{figure*}

\noindent Pathological degeneration of RGCs observed in RNFL, GC-IPL, and GCC thickness profiles can objectively monitor the progression of glaucomatous severity. However, manual extraction of these regions is a subjective and time-consuming task. Several automated layers extraction methods have been proposed in the literature to address this shortcoming \cite{Bagci2007LSSAW, Gao2015PLOSONE, Mayer2010BOE, Wang2019BOE, Sripad2018BOE}. But, to the best of our knowledge, there is no clinically validated framework that utilizes the degraded RGC profiles to diagnose and grade glaucomatous progression using ONH SD-OCT scans. Moreover, as the ONH SD-OCT scans are considered to be more significant for detecting glaucoma progression \cite{Almobarak2014IOVS}, validating an automated framework on a publicly available standardized ONH SD-OCT dataset adds significant value to the body of knowledge. 

\noindent In this paper, we present a fully automated diagnosis and grading of glaucoma from ONH SD-OCT images by analyzing pathological variations of RNFL, GC-IPL, and GCC regions. The proposed framework is unique as it employs a hybrid convolutional network for the RGC-aware diagnosis and grading of glaucoma, and it has been clinically validated with four expert clinicians. The main features of this paper are:
\begin{itemize}[leftmargin=*]
    \item A novel strategy for the classification and grading of glaucomatous progression by analyzing RNFL, GC-IPL, and GCC regions from ONH SD-OCT scans.
    \item A significantly improved and lightweight hybrid retinal analysis and grading network (RAG-Net\textsubscript{v2}) for the simultaneous pixel-level segmentation of retinal regions and scan-level classification of glaucomatous pathologies.
    \item Rigorous clinical validation of the proposed framework with four expert ophthalmologists to screen, track, and grade glaucomatous progression from high-quality ONH SD-OCT scans. The complete dataset and the annotations from the expert observers are publicly available at \url{https://data.mendeley.com/datasets/2rnnz5nz74/2}.
\end{itemize}
\noindent The rest of the paper is organized as follows. Section \ref{sec:method} describes the proposed method. Section  \ref{sec:expsetup} showcases the experimental setup. Section \ref{sec:results} presents the results, followed by detailed discussion and concluding remarks about the proposed framework in Section  \ref{sec:discussion}.  

\section{Proposed Method}\label{sec:method}
\noindent We present a novel framework that gives an RGC-aware diagnosis of glaucoma using ONH SD-OCT scans. Furthermore, it measures the severity of glaucomatous progression by analyzing the RNFL, GC-IPL, and GCC thickness profiles. The block diagram of the proposed framework is shown in Figure \ref{fig:fig2}. First of all, the input scan is preprocessed to retain the retinal area. Afterward, the preprocessed scan (containing only the retina and the ONH) is passed to the hybrid convolutional network that extracts the RNFL, GC-IPL, and GCC regions, and screens the scan against glaucoma. The thickness profiles of these extracted regions are computed and their mean values are passed as a feature vector to the supervised support vector machines (SVM) for grading the screened glaucomatous scan as either early suspect or a severe case. The detailed description of each block is presented below:

\subsection{Preprocessing}
\begin{figure}[htb]
    \includegraphics[width=1\linewidth]{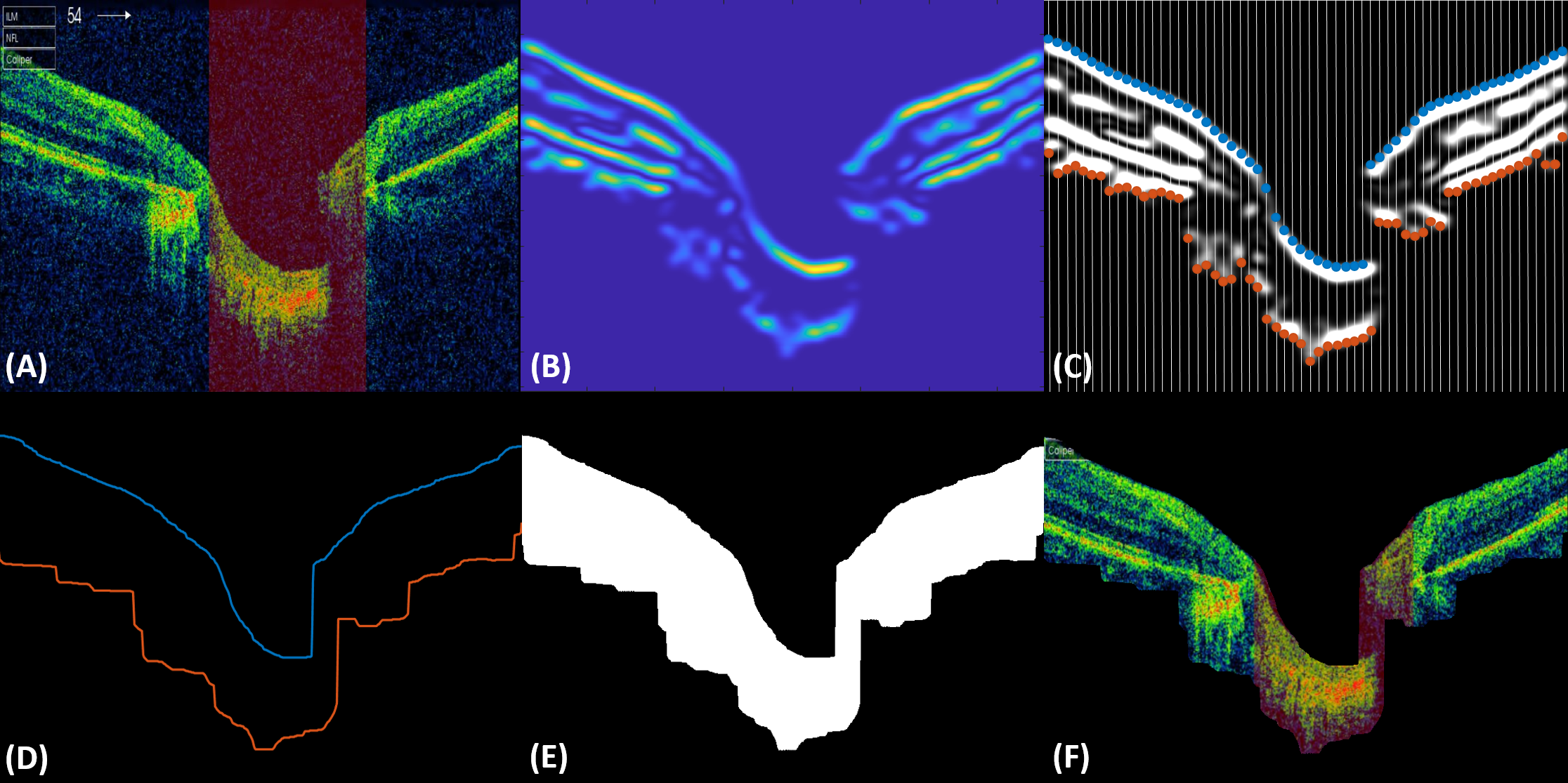}
\caption{\small Preprocessing stage (A) original ONH SD-OCT scan, (B) tensor with maximum coherency, (C) grayscale tensor from which the retinal and choroidal layer points are iteratively picked, (D) extracted ILM and choroidal layers, (E) retinal extraction mask, (F) extracted retina and ONH. }
\centering
\label{fig:fig3}
\end{figure}

\noindent  The purpose of the preprocessing is to remove the background artifacts and noisy content to obtain accurate extraction of RNFL, GC-IPL, and GCC regions. The preprocessing is performed through structure tensor \cite{st2}, which highlights the predominant orientations of the image gradients within the specified neighborhood of a pixel. For each pixel of the input image, we get a symmetric $2 \times 2$ matrix $S$ defined by the outer products of image gradients:
\begin{equation}
\small
S=\begin{bmatrix}
 \varphi * (\nabla X . \nabla X) & \varphi * (\nabla X . \nabla Y) \\ 
  \varphi * (\nabla Y . \nabla X) &  \varphi * (\nabla Y . \nabla Y)
\end{bmatrix}
\label{eq:Eq1}
\end{equation}
where the image gradients $\nabla X$ and $\nabla Y$ are oriented at $0^\circ$ and $90^\circ$, respectively. $\varphi$ denotes the parametric smoothing filter (typically a Gaussian). Because of the symmetry, three out of the four matrix elements are unique. Computing $S$ for each pixel we obtain three unique tensors from which we select the one having the maximum coherency according to their norm.  Afterward, the selected tensor is transformed as an 8-bit grayscale image. Then, the ILM and choroidal boundaries are extracted from it by detecting the first and last foreground-background transitions in each column of the scan.
To avoid outliers, we constrain the distance between consecutive pixels in the ILM and choroidal boundaries to be below a threshold $\tau$ determined empirically. 
Apart from this, the missing values in each layer are estimated through linear interpolation and are smoothed through median filtering. Then, a retinal mask is generated which is multiplied with the original scan to isolate the retinal and ONH regions as shown in Figure \ref{fig:fig3}. The complete pseudo-code to extract the retina and ONH region is presented in Algorithm \ref{algo}:

\subsection{Hybrid Convolution Framework}
\noindent We propose a  hybrid  convolutional  network (HCN)
for extracting the retinal regions and also for the classification of candidate scan as normal or glaucomatous. Using an HCN rather than a conventional classification model in this study aims to get an RGC-aware diagnosis of glaucoma. The HCN model proposed here is an improved version of the  Retinal Analysis and Grading Network (RAG-Net) \cite{Hassan2020JBHI}. The RAG-Net and its improved version will be described next.

%%%%%%%%%%%%%%%%%%%%%%%%%%%%%%%%%% Algorithm  %%%%%%%%%%%%%%%%%%%%%%%%%%%%%%%
\begin{algorithm}
\SetAlgoLined
\DontPrintSemicolon
\textbf{Input: } OCT Image $I$

\textbf{Output: } Preprocessed Image $I_{ONH}$

ILM $\gets$ $\phi$

Choroid $\gets$ $\phi$

$\tau$ $\gets$ 20

$v_1$ $\gets$ $\phi$

$v_2$ $\gets$ $\phi$

$S$ $\gets$ ComputeStructureTensor(\textit{I})

$S_u$ $\gets$ GetUniqueTensors($S$)

$S_u^t$ $\gets$ GetCoherentTensor($S_u$)

$I_u^t$ $\gets$ ConvertTensorToImage($S_u^t$)

[nRow,nCol] $\gets$ GetSize($I_u^t$)

\For {$c$ $\gets$ 1 to nCol}{
   $p_1$ $\gets$ FindFirstTransitionInRow($I_u^t$(:,$c$))
   
   $p_2$ $\gets$ FindLastTransitionInRow($I_u^t$(:,$c$))
   
   \eIf {$c$ is 1}{
      ILM($c$) $\gets$ $p_1$
      
      Choroid($c$) $\gets$ $p_2$
      
      $v_1$ = $v_2$ = $c$}{
      \textit{isP1Valid} $\gets$ CheckDistance($p_1$,ILM($v_1$), $\tau$)
      
      \textit{isP2Valid} $\gets$ CheckDistance($p_2$,Choroid($v_2$), $\tau$)
      
        \If{isP1Valid}{
           $v_1$ $\gets$ $c$
           
           ILM($v_1$) $\gets$ $p_1$
        }
        \If{isP2Valid}{
           $v_2$ $\gets$ $c$
           
           Choroid($v_2$) $\gets$ $p_2$
        }
        }
}

ILM $\gets$ InterpolateGapsAndSmoothLayer(ILM)

Choroid $\gets$ InterpolateGapsAndSmoothLayer(Choroid)

mask $\gets$ GenerateMask($I_u^t$,ILM, Choroid)

$I_{ONH}$ $\gets$ \textit{I} * mask

 \caption{Retina and ONH Extraction}
 \label{algo}
\end{algorithm}

\subsubsection{Retinal Analysis and Grading Network}
RAG-Net is a hybrid convolutional network specifically designed to extract retinal lesions and abnormalities from the macular OCT scans and give lesion-aware grading of retinal diseases by performing simultaneous pixel-level segmentation and scan-level classification. Architecturally, it contains 112 convolution layers, 111 batch normalization layers, 102 ReLUs, 6 pooling layers, 5 lambda layers, 2 softmax, and a fully connected layer \cite{Hassan2020JBHI}. Furthermore, it contains around 62.3M parameters from which around 0.1M are non-learnable. RAG-Net showed the capacity to generalize across multiple scanner specifications for retinal lesion extraction and lesion-aware grading of retinopathy and has also been applied to the multi-modal imagery for retinal lesions extraction, achieving superior performance among its competitors  \cite{Hassan2020}. However, the original RAG-Net has been found limited in discriminating similar texture objects, like retinal boundaries, when their transitional variations are small. This is because RAG-Net possesses kernels with smaller receptive fields (field of view) that do not retain accurately the contextual information of small and similar textural regions. Although, the feature pyramid module within the RAG-Net architecture tends to overcome this limitation to some extent. But overall the performance is still capping as will be shown in the results section (Section \ref{sec:results}). Moreover, the source code of RAG-Net and its complete documentation is available at \url{http://biomisa.org/index.php/downloads/} \cite{Hassan2020JBHI}.

%%%%%%%%%%%%%%%%%%%%%%%%%%%%%%%%%
\begin{figure}[htb]
    \includegraphics[width=1\linewidth]{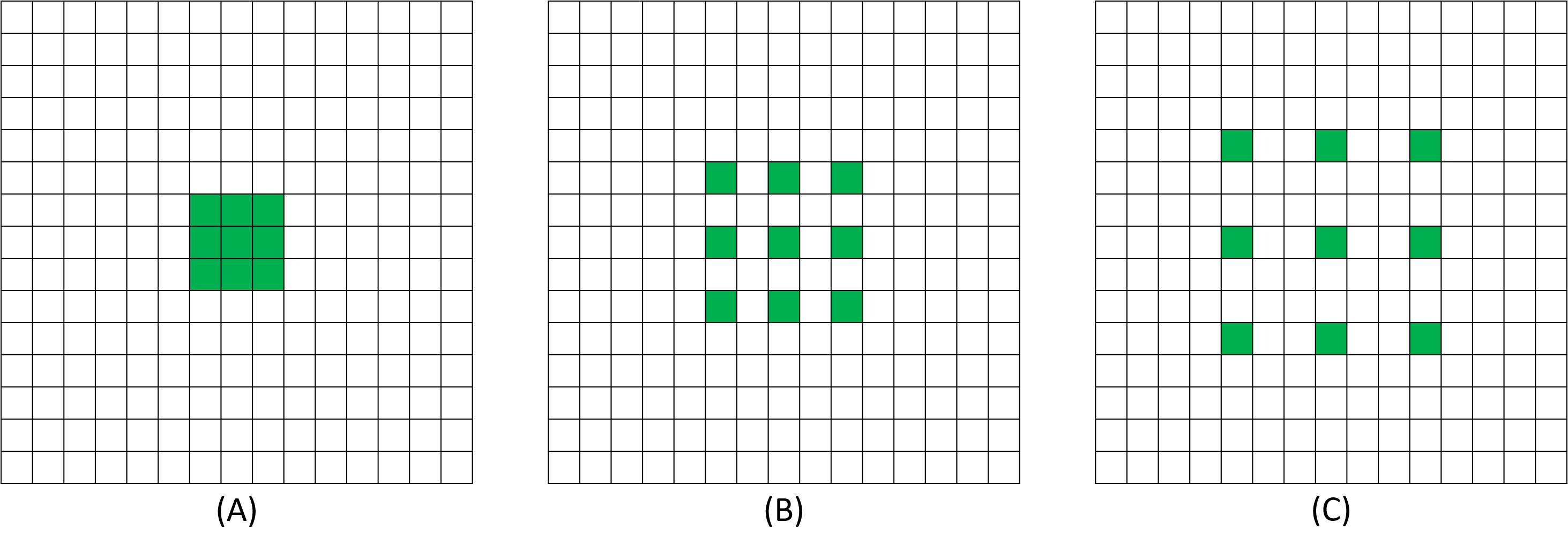}
\caption{\small Illustration of dilated convolution with $3 \times 3$ kernel and (A) dilation rate $r = 1$, (B) $r = 2$, and (C) $r = 3$. }
\centering
\label{fig:fig4}
\end{figure}
%%%%%%%%%%%%%%%%%%%%%%%%%%%%%%%%%%%%%

\begin{table}[htb]
    \centering
    \caption{RAG-Net\textsubscript{v2} hyper-parameters}
    \begin{tabular}{lcc}
         \toprule
         \textbf{Layers} & \textbf{Number of Layers} & \textbf{Parameters} \\ \hline
         Convolution & 16 & 4,847,369\\
         Pooling & 4 Average, 10 Max & 0\\
         Batch Normalization & 15 & 17,920\\
         Activation & 13 ReLU, 2 Softmax & 0\\
         Lambda & 5 & 0 \\
         Input & 2 & 0 \\
         Zero-Padding & 10 & 0 \\
         Concatenation & 1 & 0 \\
         Reshape & 1 & 0 \\
         Fully Connected and Flatten & 2 & 716,810 \\
         Classification & 1 & 22 \\ \hline
         Learnable Parameters & 5,573,161 & \\
         Non-learnable Parameters & 8,960 & \\
         Total Parameters & 5,582,121 & \\
         \bottomrule
    \end{tabular}
    \label{tab:tab1}
\end{table}

 \subsubsection{Modified Retinal Analysis and Grading Network}
\begin{figure}[htb]
    \includegraphics[width=1\linewidth]{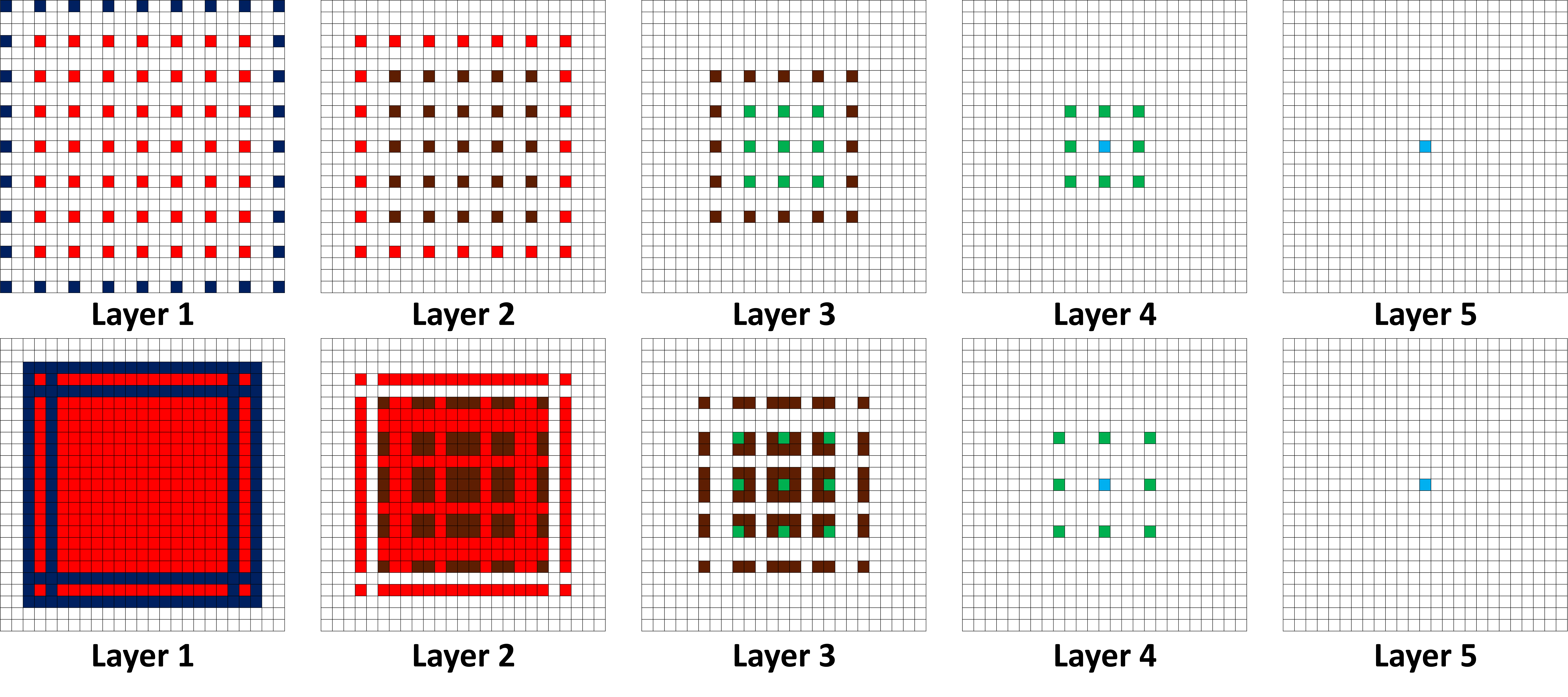}
\caption{\small A block containing 5 consecutive atrous convolution layers with fixed dilation rate ($r=3$) in top row and variable dilation rates ($r = k$ in $k^{th}$ layer where $k = 1, 2, … ,5$) in bottom row. A single pixel in layer 5 (highlighted in light blue color) is computed using (green pixels in layer 4). Similarly, the green pixels (in layer 4) are computed through brown pixels in layer 3, brown pixels are computed through red pixels in layer 2 and red pixels are computed through dark blue pixels in layer 1. It can be observed that the receptive field of a $3 \times 3$ kernel is greatly enhanced in both (top row) and (bottom row) as compared to standard linear convolutions. However, the fixed dilation rate (in the top row) introduces gridding artifacts i.e. output pixel is layer 5 is computed by totally disjoint input pixels of layer 4 and so on. Also, as the layers are cascaded, the effects of gridding artifacts can be catastrophic e.g. observe (in the top row) that how a pixel in layer 5 relates to the pixels in layer 1. Employing variable dilation rates can effectively diminish these gridding artifacts while preserving the increased field of view at the same time as shown in the bottom row.} 
\centering
\label{fig:fig5}
\end{figure}

We propose a RAG-Net modified version, dubbed RAG-Net\textsubscript{v2}, whereby the contextual-aware unit is built upon atrous convolutions (also known as dilated convolution). The atrous convolutions are formulated in a residual fashion which greatly enhances the kernels receptive field to perform more broad and context-aware filtering while maintaining the same spatial resolution \cite{Wang2018KDD}. The 2D atrous convolution is expressed as:
\begin{equation}
    g(x,y)=\sum_{i=1}^{N_1}\sum_{j=1}^{N_2} k(i,j)f(x-r*i,y-r*j)
    \label{eq:eq2}
\end{equation}
where $f$ denotes input function (typically a feature map from the previous layer), $k$ represents the $N_1 \times N_2$ kernel, $r$ is the dilation rate and $g$ denotes the convolution output (a feature map produced in the current layer). It should be noticed in the above equation that we have introduced a common dilation rate $r$ in both image dimensions to ensure a consistent reception field at both of them. When $r = 1$, then atrous convolution is simply a linear convolution as shown in Figure \ref{fig:fig4} (A), and when $r >= 1$, the receptive field is enlarged so the kernel captures wider contextual area from the input function to produce more distinctive feature maps. However, increasing the $r$ also introduces gridding artifacts \cite{Wang2018KDD} due to large gaps between convolving pixels in the input function, causing a cascading effect in the consecutive convolution layers which may result in a significant decrease of network performance. The gridding artifacts are also illustrated in Figure \ref{fig:fig5} (top row) for the stacked convolution layers.
%%%%%%%%%%%%%%%%%%%%%%%%%%%%%%%%%%%%%%%%%%%%%%%
%%%%%%%%%%%%%%%%%%%%%%%%%%%%%%%%%%%%%%
In order to compensate this, we perform atrous convolution with variable dilation factors \cite{Wang2018WACV} in the RAG-Net\textsubscript{v2} architecture. For a block of $n$ consecutive convolution layers within the network having the dilation rate ‘$r$’ where $n > r$, the dilation factors in the proposed framework are generated through $round(r-\frac{n}{2}+i)$ where $i$ varies from $0$ to $n-1$. For example, for a block containing 5 cascaded convolution layers having the dilation rate $r = 3$, the dilation factors will be [1, 2, 3, 4, 5], meaning that the first convolution layer within the block will perform standard convolution (as $r = 1$), the second layer will have $r = 2$ and so on as shown in Figure \ref{fig:fig5} (bottom row). 
Similarly, for $n = 3$, $r = 3$, the dilation factors will be [2, 3, 4]. 
The second major benefit of RAG-Net\textsubscript{v2} is that it is extremely lightweight and contains 91.04\% fewer parameters than original RAG-Net architecture (having 62,352,188 parameters in total) while achieving the better segmentation and classification performance. The detailed architectural description and hyper-parameters of RAG-Net\textsubscript{v2} are reported in Table \ref{tab:tab1} from which we can see that it contains 5,573,161 learnable and 8,960 non-learnable parameters. Moreover, rather than training RAG-Net\textsubscript{v2} from scratch, it is fine-tuned on RAG-Net weights (which are already adjusted in \cite{Hassan2020JBHI} for lesion-aware retinal image analysis) to achieve faster convergence. 

\subsection{Estimation of Retinal Profiles}
\noindent The severity of glaucoma can be effectively graded by analyzing the RGC atrophy. RGCs primarily exists within the GCC region that consists of RNFL, the ganglion cellular layer (GCL), and the inner plexiform layer (IPL) \cite{Maetschke2019PLOSONE}. To objectively evaluate the glaucoma progression, the proposed system computes the RNFL, GC-IPL, and the GCC thickness profiles. The RNFL thickness is computed by taking the absolute difference between the ILM and the GCL. Moreover, GC-IPL thickness is computed by taking the absolute difference between GCL and IPL. Also, the GCC thickness is computed by taking the absolute difference between ILM and IPL. Afterward, the mean RNFL, GC-IPL, and GCC thickness values are computed from the extracted thickness profiles which are then passed to the supervised SVM model for grading the glaucomatous progression. The reason for choosing these thickness values as features for the SVM is because they reflect the pathological degeneration of RGCs which can be used for grading glaucoma (predicted through the RAG-Net\textsubscript{v2} classification unit) as early or advanced (more severe). 

\subsection{Modified Dice-Entropy Loss Function}
\noindent In order to effectively extract the retinal regions and use them for the RGC-aware classification of glaucomatous scans, RAG-Net\textsubscript{v2} jointly optimized the dice-entropy loss which is a linear combination of both dice and cross-entropy loss as expressed below:

\begin{eqnarray}
    L_{de} & = & \alpha_1 L_d + \alpha_2 L_e
    \label{eq:eq3} \\
    L_d & = & \frac{1}{N} \sum_{i=1}^N \left(1 - \frac{2 \sum_{j=1}^C t_{i,j} p_{i,j}} {\sum_{j=1}^C t_{i,j}^2 + \sum_{j=1}^C p_{i,j}^2}\right)
    \label{eq:eq4} \\
    L_e & = & -\frac{1}{N} \sum_{i=1}^{N} \sum_{j=1}^{C} t_{i,j} \log⁡(p_{i,j})
    \label{eq:eq5}
\end{eqnarray}

\noindent where $L_d$ denotes the dice loss, $L_e$ represents the multi-category cross-entropy loss, $t_{i,j}$ represents the true labels for $i^{th}$ sample belonging to $j^{th}$ class, $p_{i,j}$ denotes the predicted probability for the $i^{th}$ sample belonging to $j^{th}$ class, $\alpha_{1,2}$ represent the loss  functions weights, $N$ represents the total number of samples in a batch, and $C$ represents the total number of classes. The dice-entropy loss penalizes RAG-Net\textsubscript{v2} to accurately segment the RNFL and GC-IPL pixels from the other retinal pixels and at the same time enables RAG-Net\textsubscript{v2} to robustly classify the ONH SD-OCT scans as healthy or glaucomatous.

\section{Experimental Setup}  \label{sec:expsetup}
\noindent This section presents a detailed description of the dataset and the training protocol. It also presents the evaluation metrics which have been used to validate the performance of the proposed framework.

\subsection{AFIO Dataset}
\noindent Armed Forces Institute of Ophthalmology (AFIO) dataset, firstly introduced in \cite{Raja2020DIB}, is a publicly available repository containing high-resolution ONH SD-OCT scans of healthy and glaucomatous subjects. The dataset has been acquired from AFIO Hospital, Rawalpindi, Pakistan. To the best of our knowledge, it is the only dataset that contains OD centered fundus and ONH centered SD-OCT scans for each subject along with the detailed cup-to-disc markings and annotations from four expert ophthalmologists. The scans within the AFIO dataset are acquired using Topcon 3D OCT-1000 sampled over four years. Furthermore, all the scans within AFIO datasets have been thoroughly graded by four expert ophthalmologists in a blind-manner (i.e. each grader does not know the grading done by his/her colleagues). Also, all four clinicians were very senior (having 20 to 25 years of professional experience in clinical ophthalmology). Moreover, the detailed specifications of the AFIO dataset are presented in Table \ref{tab:tab2}. 

\begin{table}[htb]
    \centering
    \caption{AFIO Dataset Specifications}
    \begin{tabular}{ll}
    \toprule
    Acquisition Machine & Topcon 3D OCT 1000 \\\hline
    Scan Reference & Optic Nerve Head (ONH) Centered \\
    Examination	& Dilated Pupil with Ø4.0mm (45º) Diameter\\
    Images & 196 ONH SD-OCT Images\\
    Scan Type & B-scan\\
    Resolution & 951x456\\
    Subjects & 101\\
    Categories & Healthy: 50\\
     & Glaucoma: 146\\
    \bottomrule
    \end{tabular}
    \label{tab:tab2}
\end{table}

\subsection{Training Details}
\noindent RAG-Net\textsubscript{v2} in the proposed framework is implemented using Keras APIs on Anaconda Python 3.7.4 platform\footnote{The source code is available at \url{https://github.com/taimurhassan/rag-net-v2}.}. The training is conducted for 40 epochs where each epoch lasted for 512 iterations on a machine with Intel Core i7-9750H@2.6 GHz processor and 32 GB RAM with a single NVIDIA RTX 2080 Max-Q GPU having cuDNN v7.5 and a CUDA Toolkit 10.1.243. The optimization during the training is performed through ADADELTA \cite{Zeiler2012ADADELTA} having a default learning rate of one with the decay factor of 0.95. Moreover, 70\% of the dataset is used for the training and the rest of 30\% were used for testing as per the dataset standard \cite{Raja2020DIB}. To compensate for the low number of training scans within the AFIO dataset, we fine-tuned the weights of the original RAG-Net architecture (obtained after training on more than 0.1 million macular OCT scans \cite{Hassan2020JBHI}) and also performed augmentation of the training scans. The data augmentation we performed is as follows: first, all the scans were horizontally flipped, and then they were rotated between -5 to 5 degrees. Then, we added a zero-mean white Gaussian noise with 0.01 variance. The augmentation procedure resulted in a total of 6,028 training scans to fulfill the training requirements. 

\subsection{Evaluation Metrics}
\noindent The proposed framework has been evaluated using a number of metrics described below:

\subsubsection{Confusion Matrix}
The  performance  of  the  proposed framework  for  accurately  classifying  and  grading  glaucomatous  subject  is  measured  through the confusion matrix and its associated metrics such as accuracy ($ A_{C} = \frac{T_P + T_N}{T_P + T_N + F_P + F_N}$), recall ($T_{PR} = \frac{T_P}{T_P + F_N}$), specificity ($T_{NR} = \frac{T_N}{T_N + F_P}$), false positive rate ($F_{PR} = \frac{F_P}{T_N + F_P}$), precision ($P_{PV} = \frac{T_P}{T_P + F_P}$) and F\textsubscript{1} score ($F_1 = \frac{2 * P_{PV} * T_{PR}}{P_{PV} + T_{PR}}$), where $T_P$ denotes true positives, $T_N$ denotes the true negatives, $F_P$ denotes the false positives, and $F_N$ denotes the false negatives. To measure classification performance, $T_P$, $F_P$, $T_N$ and $F_N$ are calculated scan-wise.

\subsubsection{Receiver Operating Characteristics (ROC) Curve}
ROC curve indicates the capacity of the proposed framework to correctly classify and grade healthy and glaucomatous pathologies at various classification thresholds. Moreover, the performance through ROC curves is quantitatively measured through $AUC$ scores.

\subsubsection{Dice Coefficient ($D_C$)}
The dice coefficient ($D_C$) measures how well the proposed framework segments the RNFL, GC-IPL, and GCC regions as compared to their ground truths, and it is computed through: ($D_C= \frac{2 * TP}{2 * TP + FN + FP}$). Here $T_P$, $F_P$, and $F_N$ are calculated pixel-wise where $T_P$ indicates the correct extraction of positives (RNFL, GC-IPL, and GCC regions), $F_P$ indicates the misclassified background pixels, and $F_N$ denotes those positive pixels which have been missed by the proposed framework. Afterward, the mean dice coefficient ($\mu_{DC}$) is computed by taking an average of $D_C$ scores scan-wise across the whole dataset.

\subsubsection{Mask Precision}
To further validate the performance of the proposed framework for extracting RNFL, GC-IPL, and GCC regions, we used the mask precision ($m_p$) metric. Unlike the dice coefficient, $m_p$ measures both the capacity of the proposed framework in accurately recognizing the RNFL, GC-IPL, and GCC regions as well as extracting their corresponding masks. First of all, the dice coefficient  $D_C$ of the extracted regions is computed in each image using their ground truths. If $D_C \geq 0.5$, then the $m_p$ (for each region) is computed pixel-wise through $m_{p} = \frac{T_P}{T_P + F_P}$. However, if the dice coefficient is below 0.5, then the whole region is considered as $F_P$, resulting in a $m_p$ score of 0. Moreover, the mean mask precision $\mu_{mp}$ is computed by taking the average of $m_p$ for each region as $\mu_{mp}=\sum_{k=0}^{c} m_p(k)$, where $c$ denotes the number of classes (regions).

\subsubsection{Clinical Validation}
Apart from using performance metrics, we clinically validated the glaucomatous screening and grading performance of the proposed framework with four expert ophthalmologists using the standardized Pearson correlation coefficient ($r_c$) and its statistical significance measured through the $p$-value.

\section{Results} \label{sec:results}
\noindent The proposed framework has been thoroughly evaluated on the AFIO dataset for the RGC-aware diagnosis and grading of glaucoma. First, we present an ablative analysis to evaluate different segmentation models for the extraction of RNFL, GC-IPL, and GCC regions. Afterward, we present a detailed comparison of the proposed framework with state-of-the-art solutions for extracting the RGC regions as well as screening glaucomatous subjects. Apart from this, we also present the clinical validation of our RAG-Net\textsubscript{v2} driven grading system with four expert ophthalmologists.

\subsection{Ablation Study}
\begin{figure*}[t]
    \includegraphics[width=1\linewidth]{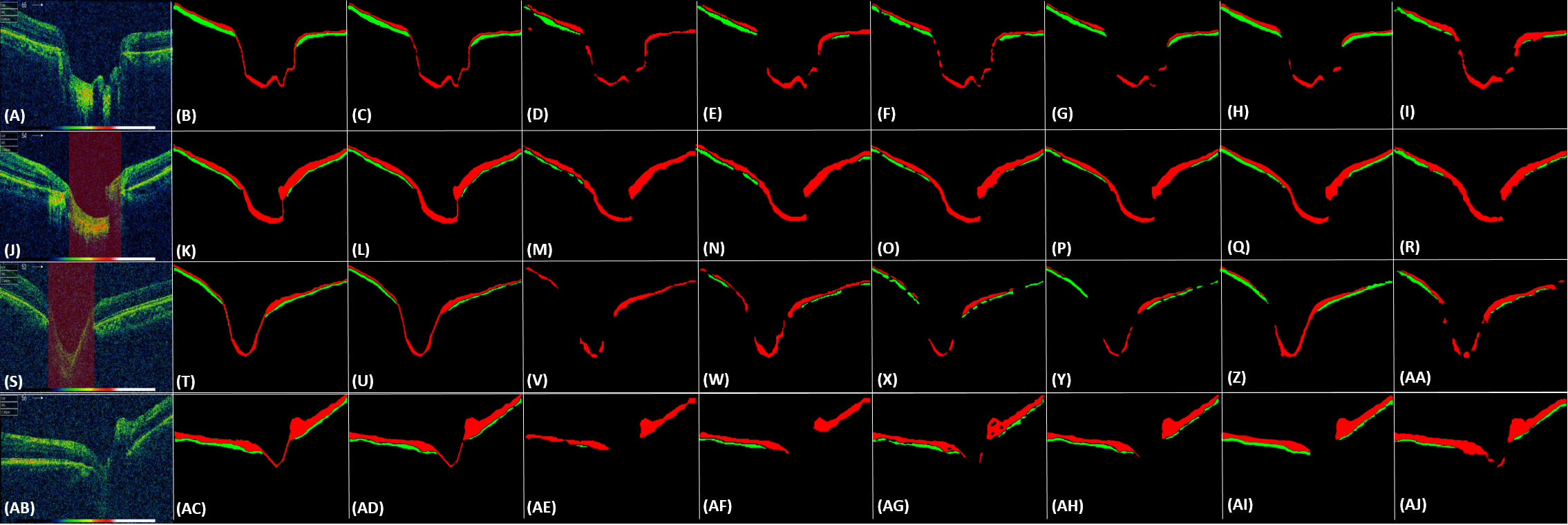}
\caption{\small Performance comparison of deep segmentation models for the extraction of RNFL (shown in red color) and GC-IPL regions (shown in green color). Left to right, column-wise: Original scans, ground truths, the performance of RAG-Net\textsubscript{v2}, RAG-Net, PSPNet, SegNet, UNet, FCN-8, and FCN-32.} 
\centering
\label{fig:fig6}
\end{figure*}

\noindent The ablative aspect of this research involves choosing the segmentation framework that can accurately extract the retinal regions such as the RNFL and the GC-IPL regions to compute the GCC profiles and grade glaucomatous subjects accordingly. For this purpose, we compared the performance of the proposed RAG-Net\textsubscript{v2} segmentation unit with the popular state-of-the-art PSPNet \cite{PSPNet}, SegNet \cite{SegNet}, UNet \cite{unet}, FCN-(8, 32) \cite{fcn}, as well as  our original RAG-Net architecture \cite{Hassan2020JBHI}. The extraction performance is shown in Table \ref{tab:tab3}, where we can observe that RAG-Net\textsubscript{v2} achieved the overall best $\mu_{DC}$ score of 0.8697 leading the second-best FCN-8 \cite{fcn} by 2.78\%. Moreover, the performance of PSPNet \cite{PSPNet} and FCN-8 \cite{fcn} are extremely comparable  as  the FCN-8 \cite{fcn} is leading PSPNet \cite{PSPNet} by 0.035\% only. Similarly, FCN-8 \cite{fcn} is leading RAG-Net \cite{Hassan2020JBHI}  by 7.12\%. If we look at the performance of each model for extracting individual regions, we can see that the best score for extracting RNFL is achieved by FCN-8 \cite{fcn}, though it’s leading PSPNet \cite{PSPNet} by 0.011\% and RAG-Net\textsubscript{v2} by 0.651\%. For extracting GC-IPL, the best performance is achieved by RAG-Net\textsubscript{v2}, leading the second-best FCN-8 \cite{fcn} by 6.28\%. The best performance for extracting GCC regions is also achieved by the RAG-Net\textsubscript{v2} i.e. $\mu_{DC}$ score of 0.8698. In terms of $\mu_{mp}$, the overall best performance is also achieved by RAG-Net\textsubscript{v2} as shown in Table \ref{tab:tab4}, leading the second-best FCN-8 by 1.78\%.  It leads the second-best FCN-8 by 3.19\% and 3.38\%, respectively for extracting GC-IPL and GCC regions. However, for the RNFL extraction, it lags from both FCN-8 and PSPNet by 1.05\% and 0.76\%. But overall, RAG-Net\textsubscript{v2} outperforms FCN-8 and PSPNet with a larger margin in extracting RNFL, GC-IPL, and GCC regions. Figure \ref{fig:fig6} showcases some qualitative comparison of all the hybrid and conventional segmentation models. Here, scan (A), (J), and (S) depict glaucomatous pathologies whereas scan (AB) depicts a healthy pathology. Both RAG-Net\textsubscript{v2} and FCN-8 produced good performance (comparing it with the ground truths) in extracting the RNFL and GC-IPL regions.
But RAG-Net\textsubscript{v2} has an upper hand in extracting GC-IPL boundaries from both healthy and glaucomatous scans (highlighted in green color). Furthermore, the original RAG-Net \cite{Hassan2020JBHI} was found limited in accurately extract RGC regions especially in scans (D), (M), (V), and (AE). Such limitation is because RAG-Net \cite{Hassan2020JBHI} cannot well differentiate similar textural patterns (often depicted in the retinal layers and boundaries). Moreover, because RAG-Net\textsubscript{v2} achieves the overall best performance for extracting RGC regions (as evident from Table \ref{tab:tab3}, Table \ref{tab:tab4} and Figure \ref{fig:fig6}), besides having the capacity to classify and grade glaucomatous pathologies, we chose it in the proposed framework for further analysis.

\subsection{Extraction of RNFL, GC-IPL and GCC Regions}
\noindent To the best of our knowledge, all the existing methods (except \cite{Maetschke2019PLOSONE}) which have been proposed for the extraction of RNFL, GC-IPL and GCC regions have been validated on either local in-house datasets or on publicly available datasets which only contains macular OCT scans. Moreover, the framework proposed in \cite{Maetschke2019PLOSONE} is designed for screening normal and glaucomatous pathologies without paying attention to RGC atrophy. The comparison of RAG-Net\textsubscript{v2} with state-of-the-art literature \cite{Sripad2018BOE, Wang2019BOE} for the extraction of RNFL, GC-IPL and GCC regions is indirect as the experimental protocols and the datasets were different (see Table \ref{tab:tab9}). Nonetheless, this indirect comparison highlights the capacity of RAG-Net\textsubscript{v2} for extracting the retinal regions while simultaneously diagnosing and grading the glaucomatous pathologies as compared to competitive methods. As reported in Table \ref{tab:tab9}, our proposed framework lags from DRUNET \cite{Sripad2018BOE} by 5.49\% in terms of $\mu_{DC}$. However, their dataset is not public and its size is almost the half of AFIO dataset. Furthermore, DRUNET is only a conventional segmentation model that does not possess glaucomatous screening and grading capabilities as compared to RAG-Net\textsubscript{v2}. The joint segmentation and classification pipeline, termed bi-decision \cite{Wang2019BOE} achieved the mean dice coefficient of 0.72 for extracting the retinal regions. 
However, for glaucoma screening, bi-decision \cite{Wang2019BOE} is only validated on the local dataset (although the authors verified bi-decision on the selective 110 DME affected scans as well from the publicly available Duke dataset  \cite{Chiu2015BOE}). Here, we want to highlight that we have already validated the original RAG-Net \cite{Hassan2020JBHI} on 43,613 macular OCT scans from five publicly available datasets (including the Duke dataset \cite{Chiu2015BOE}), and here our emphasis is on extending our hybrid framework for the RGC-aware diagnosis and grading of glaucoma using high-quality ONH SD-OCT scans from the publicly available dataset, and for reproducing the clinical trials.  

\begin{table}[htb]
    \centering
    \caption{Comparison of the proposed framework with DRUNET \cite{Sripad2018BOE} and Bi-decision \cite{Wang2019BOE}. The abbreviations are DS: Dataset Size, DPA: Dataset Publicly Available, EP: Experimentation Protocol, GS: Glaucomatous Screening, GG: Glaucomatous Grading, TR: Training, TE: Testing, CV: Cross-Validation.}
    \begin{tabular}{cccc}
        \toprule
         & Proposed & DRUNET \cite{Sripad2018BOE} & Bi-decision \cite{Wang2019BOE}\\\hline
        DS & 196\textsuperscript{\#} (ONH) & 100\textsuperscript{\#} (ONH) & 1,114* \\
        DPA & Yes & No & No \\
        EP & TR: 137, TE: 59 & TR: 40, TE: 60 & 3-Fold CV\\
        GS & Yes & No & Yes \\
        GG & Yes & No & No \\
        $\mu_{DC}$ & 0.86 & 0.91 & 0.72 \\ 
        \bottomrule
    \end{tabular}
    \\ \justify * 1,004 are circular OCT scans from the local dataset, and 110 selected macular scans are taken from Duke dataset \cite{Chiu2015BOE} representing DME pathologies.
    \\ \textsuperscript{\#} This count represents the dataset size excluding the augmented scans. 
    
    \label{tab:tab9}
\end{table}

\begin{table}[htb]
    \centering
    \caption{Comparison of segmentation models in terms of $\mu_{DC}$ for extracting the RNFL, GC-IPL, and GCC regions.} 
    \begin{tabular}{ccccc}
    \toprule
         Framework	&RNFL	&GC-IPL	&GCC &Mean\\ \hline
         RAG-Net\textsubscript{v2} & 0.8692 & \textbf{0.8703} & \textbf{0.8698} & \textbf{0.8697} \\
         RAG-Net \cite{Hassan2020JBHI}  & 0.8192 & 0.7508 & 0.7860 & 0.7853\\
         PSPNet \cite{PSPNet}  & 0.8748 & 0.8151 & 0.8457 & 0.8452 \\
         SegNet \cite{SegNet}  & 0.8111 & 0.6945 & 0.7555 & 0.7537 \\
         UNet \cite{unet}  & 0.8216 & 0.8253 & 0.8234 & 0.8234 \\
         FCN-32 \cite{fcn}  & 0.8638 & 0.7470 & 0.8083 & 0.8064\\
         FCN-8 \cite{fcn}  & \textbf{0.8749} & 0.8156 & 0.8460 & 0.8455\\
    \bottomrule
    \end{tabular}
    \label{tab:tab3}
\end{table}

\begin{table}[htb]
    \centering
    \caption{Comparison of segmentation models in terms of $\mu_{mp}$  for extracting the RNFL, GC-IPL, and GCC regions.} 
    \begin{tabular}{ccccc}
    \toprule
         Framework	&RNFL	&GC-IPL	&GCC &Mean\\ \hline
         RAG-Net\textsubscript{v2} &0.8410	& \textbf{0.8108} &	\textbf{0.7915} & \textbf{0.8144} \\	
         RAG-Net \cite{Hassan2020JBHI}  & 0.7661 & 0.6701 & 0.6016 & 0.6792\\
         PSPNet \cite{PSPNet}  & 0.8475 & 0.7685 & 0.7229 & 0.7796 \\
         SegNet \cite{SegNet}  & 0.7402 & 0.6978 & 0.6630 & 0.7003 \\
         UNet \cite{unet}  & 0.8082 & 0.7796 & 0.7291 & 0.7723 \\
         FCN-8 \cite{fcn}  & \textbf{0.8500} & 0.7849 & 0.7647 & 0.7999\\
         FCN-32 \cite{fcn}  & 0.7623 & 0.7080 & 0.6867 & 0.7190\\
    \bottomrule
    \end{tabular}
    \label{tab:tab4}
\end{table}

\subsection{Classification of Glaucomatous Scans}
\begin{figure}[htb]
    \centering
    \includegraphics[width=0.922\linewidth]{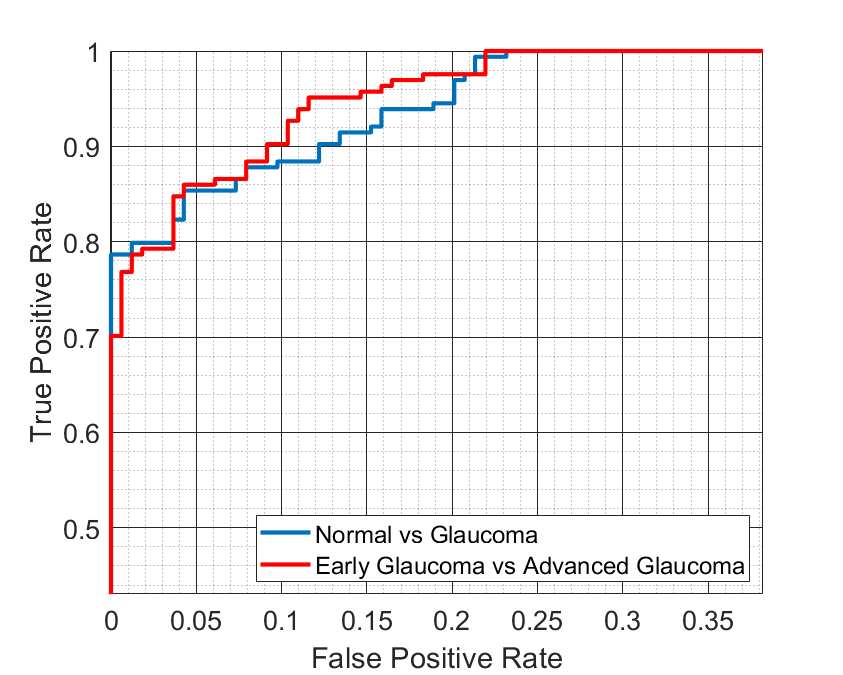}
\caption{\small ROC curve highlighting the performance of RAG-Net\textsubscript{v2} for classifying and grading glaucoma. }
\label{fig:fig7}
\end{figure}

\noindent Since the thinness of the RNFL, GC-IPL and GCC profiles highlight the degeneration of RGCs, which directly reflects glaucomatous progression. We have utilized the encoder end of the RAG-Net\textsubscript{v2} to perform RGC-aware classification of healthy and classification pathologies. After training the RAG-Net\textsubscript{v2} for extracting the RGC regions, the trained weights are used for screening healthy and glaucomatous pathologies through the RAG-Net\textsubscript{v2} classification unit. 
The  performance of RAGNet\textsubscript{v2} classification unit for screening glaucoma is measured through standard metric such as $A_C$, $T_{PR}$, $T_{NR}$, $P_{PV}$, $AUC$ and $F_1$ score.  
As reported in Table \ref{tab:tab5}, the proposed framework achieves 0.958\% better results as compared with \cite{Khalil2018Access} and \cite{Khalil2020Wiley}, and 12.5\% better results as compared to \cite{Khalil2017IET} in terms of accuracy. The comparison with \cite{Khalil2017IET} is indirect as the authors only used fundus imagery for the classification of glaucomatous pathologies\footnote{We also evaluated the RAGNet\textsubscript{v2} classification unit for screening glaucoma using fundus images. Please see the supplementary material for more details on these additional experiments.}.  
However, the comparison with \cite{Khalil2018Access} and \cite{Khalil2020Wiley} is fair and direct as both of these frameworks were tested on the AFIO dataset using the same experimental protocols. We can further observe the capacity of the proposed framework in screening glaucomatous pathologies through the $T_{PR}$ ratings in Table \ref{tab:tab5}, where the  RAG-Net\textsubscript{v2} achieves 1.73\% better performance than \cite{Khalil2018Access}, and also leading \cite{Khalil2020Wiley} by 4.41\%. This performance gain is achieved because RAG-Net\textsubscript{v2} pays attention to the pathological variations of RGCs related to the progression of glaucoma.
The classification performance of RAG-Net\textsubscript{v2} is also evaluated through ROC curve as shown in Figure \ref{fig:fig7}, and compared with \cite{Maetschke2019PLOSONE} in terms of $AUC$ as shown in Table \ref{tab:tab5}, where we can see that RAG-Net\textsubscript{v2} leads \cite{Maetschke2019PLOSONE} by 4.77\%. However, this comparison is also indirect as both frameworks are tested on different datasets.

\begin{table}[htb]
    \centering
    \caption{Comparison of classification performance of RAG-Net\textsubscript{v2} with state-of-the-art solutions. Bold indicates the best scores while the second-best scores are underlined.}
    \begin{tabular}{ccccccc}
    \toprule
         Metric	& Proposed & \cite{Khalil2018Access} & \cite{Khalil2020Wiley} & \cite{Khalil2017IET} & \cite{Maetschke2019PLOSONE} & \cite{Wang2019BOE}\\ \hline
         $A_C$ & 	\textbf{0.9491} & \underline{0.9400} & \underline{0.9400} & 0.8300 &	- & 0.8140 \\
         $T_{PR}$ & \textbf{0.9714} & \underline{0.9545} & 0.9285 & 0.8846 & - & - \\
         $T_{NR}$ & 0.9166 & \underline{0.9285} & \textbf{0.9545} & 0.7708 & - & - \\
         $F_{PR}$ & 0.0834 & \underline{0.0714} & \textbf{0.0455} & 0.2292 & - & - \\
         $P_{PV}$ & \underline{0.9444} & \textbf{0.9629} & \textbf{0.9629} & 0.8604 & - & - \\
         $F_1$ & \textbf{0.9577} & \underline{0.9453} & \underline{0.9453} & 0.8723 & - & - \\
         $AUC$ & \textbf{0.9871} & -  & - & - & \underline{0.9400} & 0.8640 \\
         \bottomrule
    \end{tabular}
    \label{tab:tab5}
\end{table}

\subsection{Profile Extraction}
\begin{figure}[htb]
    \includegraphics[width=1\linewidth]{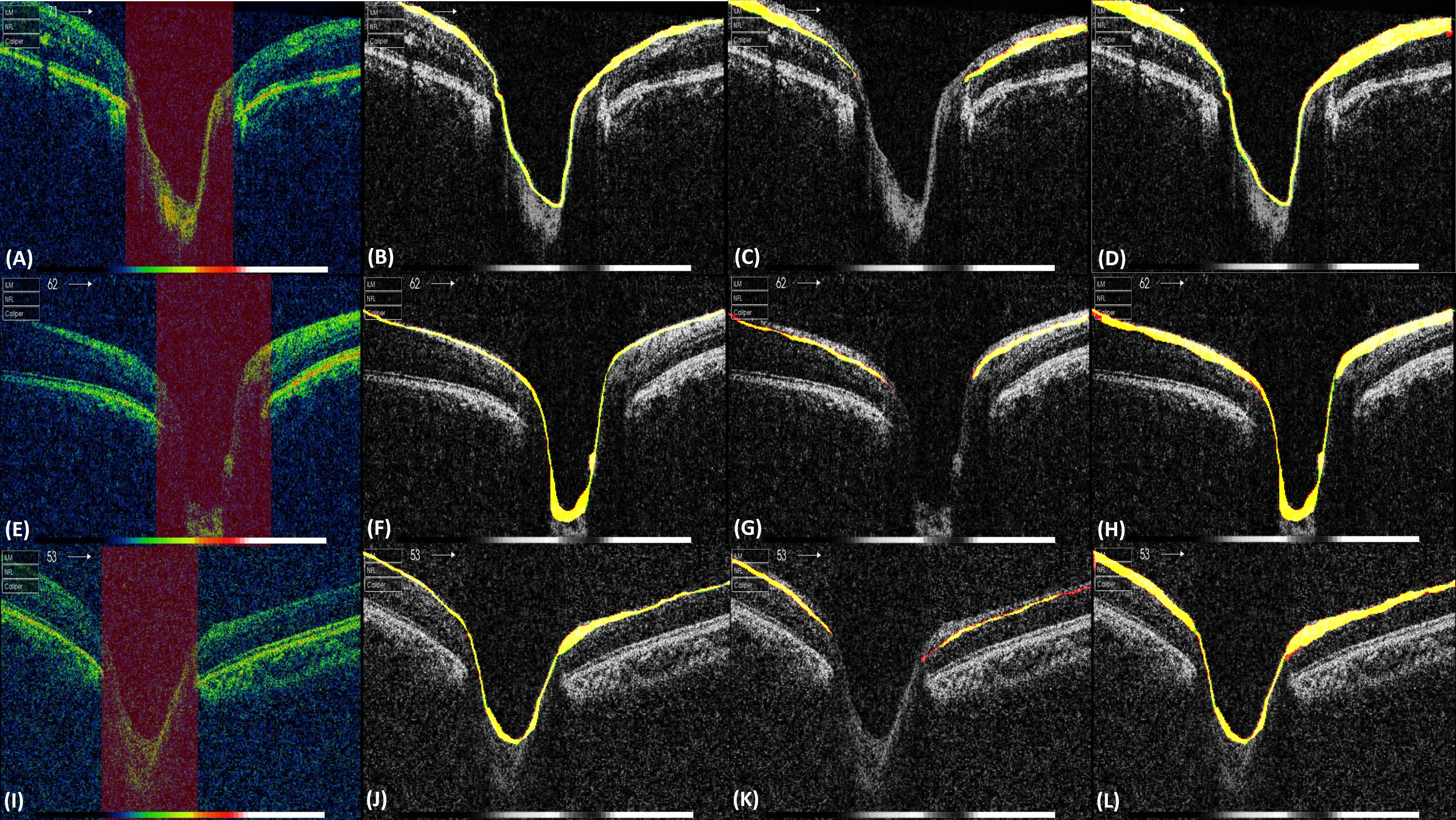}
\caption{\small Examples of normal, early and advanced stage glaucomatous scans from which the RNFL, GC-IPL, and GCC regions are extracted. The scans are intentionally made grayscale to highlight the extraction results as compared to the ground truths. The yellow color indicates complete overlap with the ground truths while other colors indicate incorrect extractions. Please zoom-in to best see the results.}
\centering
\label{fig:fig8}
\end{figure}

\noindent The novel aspect of the proposed framework is that it can grade glaucomatous progression by analyzing pathological variations of RGCs through RNFL, GC-IPL, and GCC regions represented within the ONH SD-OCT scans. Table \ref{tab:tab6} reports the mean RNFL, GC-IPL, and GCC region thickness range for the early and advanced stage glaucomatous pathologies. 
We can observe here that the RAG-Net\textsubscript{v2} achieved the mean RNFL thickness of 93.50$\mu m$ for early glaucomatous suspects and 69.46$\mu m$ for the advanced glaucomatous stage. These RNFL thickness ranges were obtained from scans of publicly available AFIO datasets and confirmed by expert clinicians. They deviate from the ranges defined by \cite{El-Naby2014Egyptian} by 2.67\% and 3.25\%, respectively, for the early and advanced stage glaucoma. We can also observe that GC-IPL and GCC profiles provide a unique distinction between glaucomatous severity, and can contribute positively towards grading it. 
In Figure \ref{fig:fig8}, we report some qualitative examples  highlighting the segmentation performance of RAG-Net\textsubscript{v2}.
The scans in this figure (except the original ones in the first column) are intentionally converted to grayscale so that the extracted regions can be visualized.
Scans (A), (E), and (I) represent normal, early glaucomatous suspect and advanced-stage glaucoma, respectively from which RAG-Net\textsubscript{v2} has accurately extracted the RNFL, GC-IPL, and GCC profiles. For example, we can see RNFL thinning in the scan (F) and (J) as compared to (B), and how precisely it is picked by RAG-Net\textsubscript{v2}. Also, the yellow color in the scans of Figure \ref{fig:fig8} (except the first column) indicates the overlap for the extracted retinal regions with the ground truth whereas other colors indicate incorrectly segmented regions (these regions are very small, please zoom in to best see them). 

\begin{table}[htb]
    \centering
    \caption{Mean RNFL, GC-IPL, and GCC thickness profiles extracted by the proposed framework for early and advanced glaucomatous pathologies.}
    \begin{tabular}{ccc}
         \toprule
         Mean Thickness & Early & Advanced\\\hline
         RNFL & 93.50$\pm$9.84 & 69.46$\pm$5.17\\
         GC-IPL & 62.23$\pm$5.67 & 33.96$\pm$7.53\\
         GCC & 155.73$\pm$13.10 & 103.42$\pm$10.27\\
         RNFL \cite{El-Naby2014Egyptian} & 91.00$\pm$7.28 & 67.20$\pm$7.06\\
        \bottomrule
    \end{tabular}
    \label{tab:tab6}
\end{table}

\begin{figure}[b]
    \centering
    \includegraphics[width=0.95\linewidth]{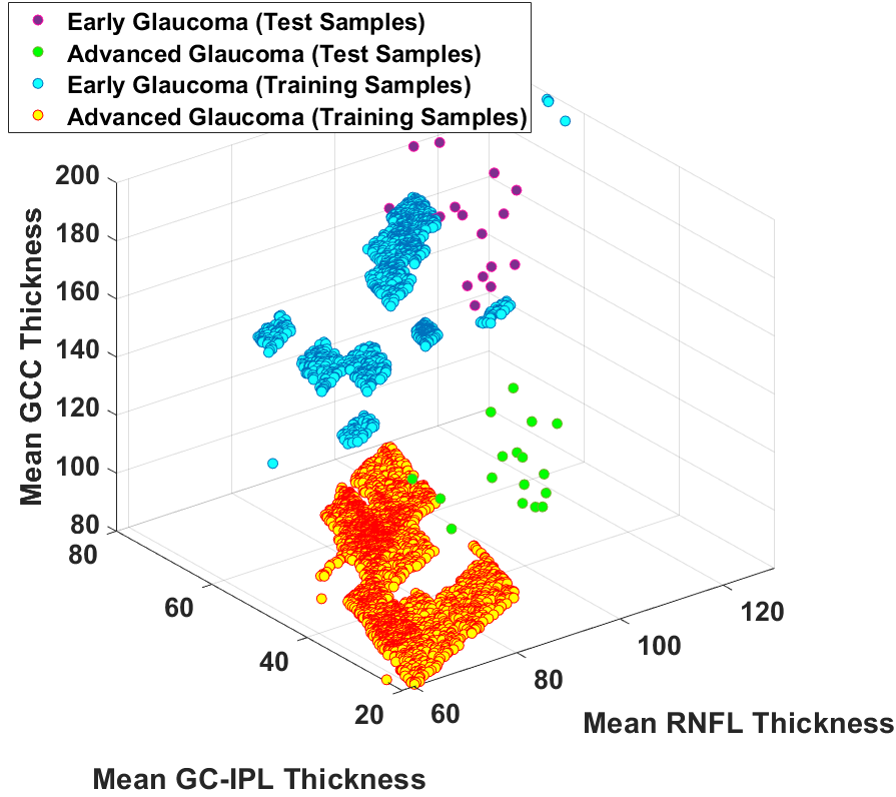}
\caption{\small Distinctive RNFL, GC-IPL, and GCC thickness profiles to discriminate early and advanced stages glaucomatous pathologies.}
\label{fig:fig9}
\end{figure}

\noindent In Figure \ref{fig:fig9}  we report plots of the class distribution for the early and advanced glaucomatous classes in the feature space defined by RNFL, GC-IPL, and GCC  profiles. We can observe that the thinning of these profiles can clearly distinguish glaucoma progression. We notice that the training samples in Figure \ref{fig:fig9} are highly clustered compared to the testing samples. This is due to the fact that these training samples are generated through data augmentation and, thus, highly correlated with the actual samples. Overall, these profiles can show clear discrimination and great potential to be used with an SVM classifier for grading glaucoma progression. 

\subsection{RGC Aware Grading of Glaucoma}
\begin{figure}[t]
    \includegraphics[width=1\linewidth]{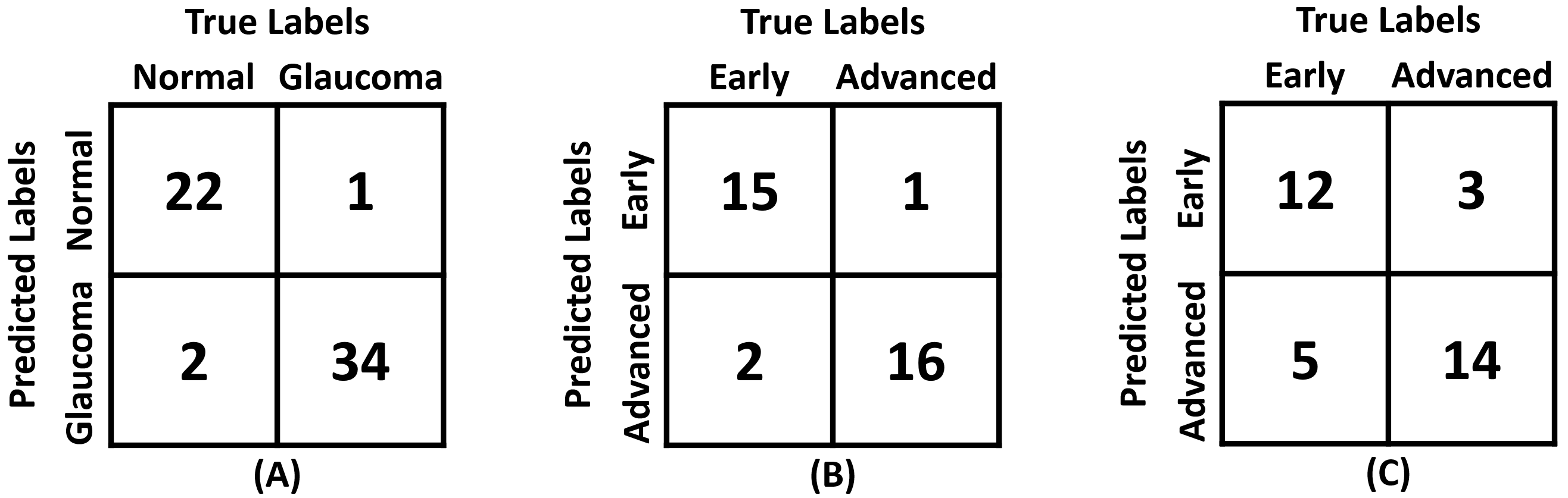}
\caption{\small Confusion matrix representing (A) classification of healthy and glaucomatous ONH SD-OCT scans, (B) grading of early suspects, and advanced-stage glaucoma, (C) glaucomatous grading using mean RNFL thickness threshold. }
\centering
\label{fig:fig10}
\end{figure}

\noindent Using the RNFL, GC-IPL, and GCC profiles as distinctive features, the proposed system provides RGC-aware grading of glaucoma using the SVM classification model. Both the classification and grading performance of the proposed framework is shown through confusion matrices in Figure \ref{fig:fig10} (A) and Figure \ref{fig:fig10} (B), respectively. Moreover, Figure \ref{fig:fig10} (C) reports the grading performance using only the mean RNFL thickness threshold (see Table \ref{tab:tab6}). The classification between normal and glaucomatous scans is performed through RAG-Net\textsubscript{v2} classification unit by directly passing the preprocessed ONH SD-OCT scans whereas the RGC-aware grading of glaucoma as early suspect or advanced stage (reported in Figure \ref{fig:fig10}-B) is performed through SVM based on RNFL, GC-IPL, and GCC profiles. In Figure \ref{fig:fig10} (B), we can see that the proposed framework achieved an accuracy of 0.9117 for screening early and advanced stage glaucoma cases which is 16.123\% superior compared to grading approach based on analyzing only the RNFL thickness. 
Also, out of 34 test samples in Figure \ref{fig:fig10} (B), three are falsely graded i.e. two early cases are graded as severe and one severe case is graded as having early glaucoma symptoms. But, these three misclassifications have less impact on the overall performance of the proposed system because both of them have been correctly classified as having glaucoma. In Figure \ref{fig:fig10} (A), we notice that one misclassification  (out of 35)  of a glaucomatous scan predicted as normal by our system. This is one challenging boundary case showing very early glaucomatous symptoms i.e. having the cup-to-disc ratio of 0.325. All the four ophthalmologists have considered this as an early suspect (please see the recommendation of the ophthalmologists in the patient record sheet within the dataset for more detail. The misclassified case has a name '149155\_20150914\_095855\_Color\_R\_001.jpg'). 

\subsection{Clinical Trials}
\noindent We have also performed a series of clinical trials in which we cross-validated the predictions of the proposed framework with the recommendations from the expert clinicians. Table \ref{tab:tab7} reports samples of the clinical trials along with the recommendations of the four expert ophthalmologists (publicly available in the dataset package \cite{Raja2020DIB}). Furthermore, we report in Table \ref{tab:tab10} the Pearson correlation analysis along with its statistical significance showcasing the clinical validation of the proposed framework with each clinician. $r_c$ ranges between -1 to +1 where -1 indicates the strong negative association between the two entities, +1 shows the strong positive association and 0 depicts that both entities are not related. Furthermore, $p$-value $<$ 0.05 indicates that the obtained $r_c$ score is statistically significant. From Table \ref{tab:tab10}, we can observe that 
although the recommendations contradict with each other (as they are marked by each clinician based on his/her own experience), the proposed framework achieves the highest correlation with Clinician-4 (i.e. $r_c$ = 0.9236), having $p$-value of 4.40 $ \times$ 10\textsuperscript{-58}). Moreover, the minimum $r_c$ score is achieved with the grading of Clinician-3 (i.e. $r_c = $ 0.6863), but it is still quite significant. In $8^{th}$ row of Table \ref{tab:tab7}, we have an exception where all clinicians have marked the scan as having early glaucomatous symptoms, and the proposed framework grades it as depicting advanced stage pathology. This indicates the proposed framework is tuned in a way to prioritize advanced stage subjects that need immediate attention and treatment to prevent vision loss and blindness. We also report in Table \ref{tab:tab7} the fundus scan with each case to help the readers (and other clinicians) in cross-verifying the predictions made by the proposed framework by correlating the cup-to-disc ratios. 
Through these fundus scans, we can also notice that the ONH SD-OCT scans graded as having early or advanced stage glaucoma typically contains a cup-to-disc ratio of 0.6 or above which is normally considered to imply glaucoma \cite{Illinois}.

\begin{table}[htb]
    \centering
    \caption{Clinical validation of the proposed framework for glaucomatous screening with respect to the recommendation from four expert ophthalmologists. C1: Clinician-1, C2: Clinician-2, C3: Clinician-3, C4: Clinician-4, PF: Proposed Framework, AG: Advanced Glaucoma, EG: Early Glaucoma, H: Healthy. Fundus scans are provided with each ONH scan for reader cross-verification.}
    \begin{tabular}{>{\centering\arraybackslash} m{2.9cm} >{\centering\arraybackslash} m{1.5cm} >{\centering\arraybackslash} m{.25cm} >{\centering\arraybackslash} m{.25cm} >{\centering\arraybackslash} m{.25cm} >{\centering\arraybackslash} m{.25cm} >{\centering\arraybackslash} m{.25cm}}
        \toprule
        \vspace{0.25cm}ONH Scan\vspace{0.25cm} & \vspace{0.25cm}Fundus\vspace{0.25cm} & \vspace{0.25cm}C1\vspace{0.25cm} & \vspace{0.25cm}C2\vspace{0.25cm} & \vspace{0.25cm}C3\vspace{0.25cm} & \vspace{0.25cm}C4\vspace{0.25cm} & \vspace{0.25cm}PF\vspace{0.25cm}\\\hline
        
        \vspace{0.15cm}\includegraphics[scale=0.0875]{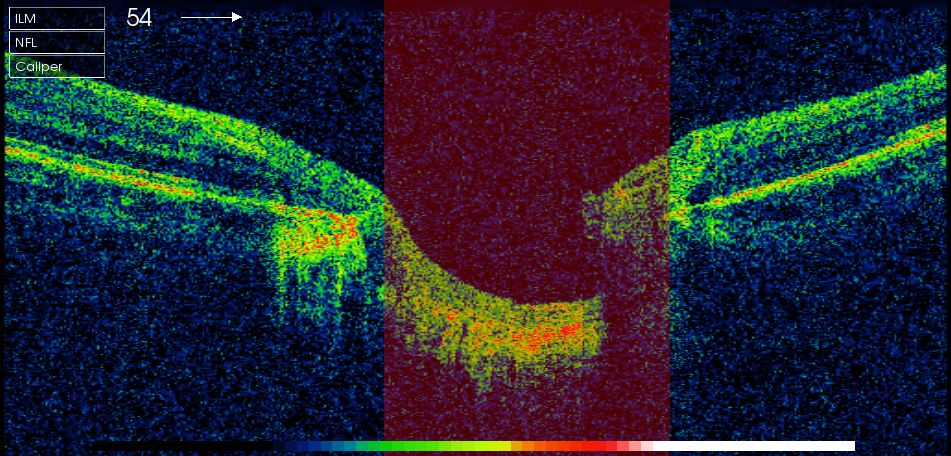} & \vspace{0.15cm}\includegraphics[scale=0.04375]{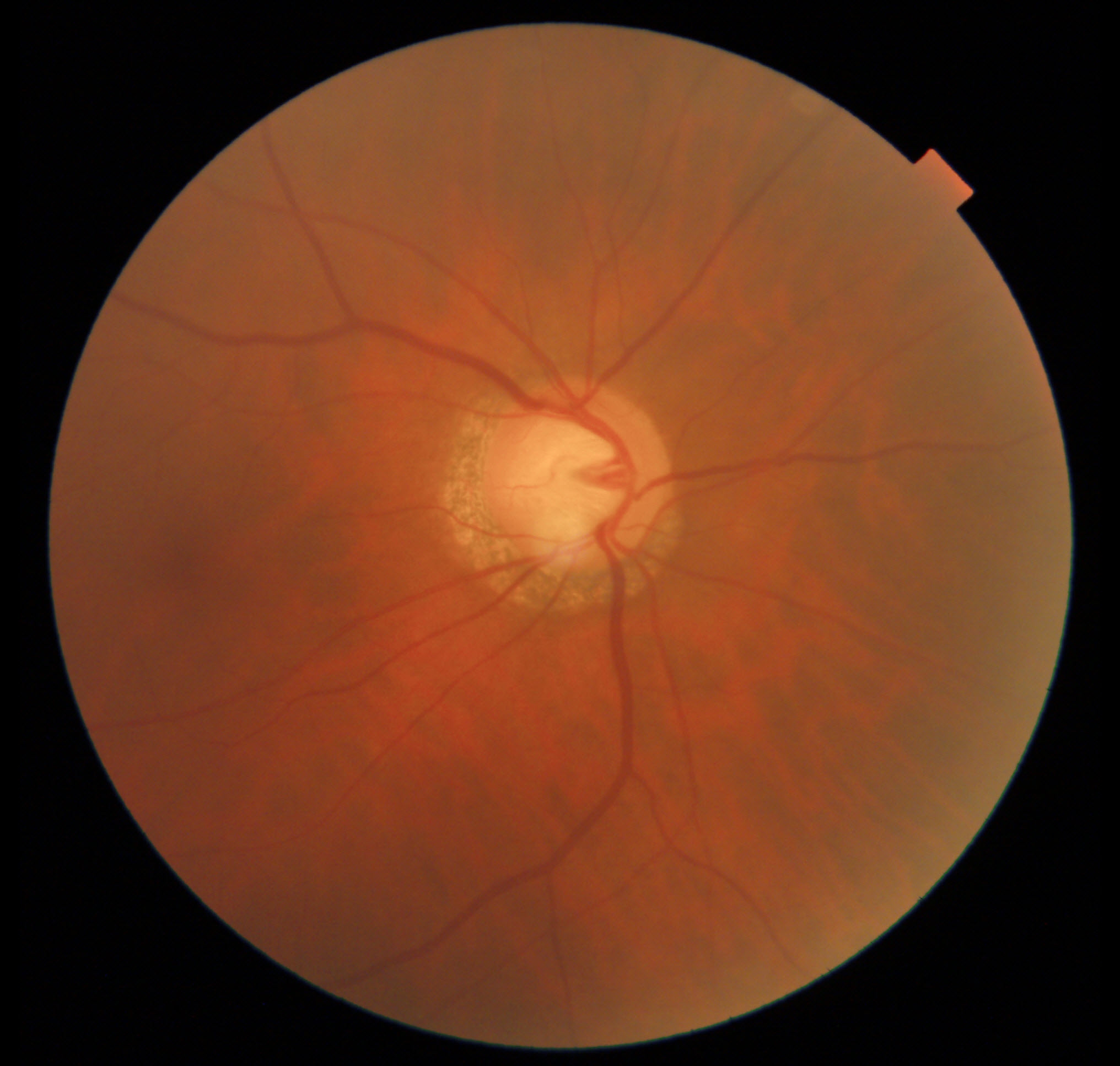} & \vspace{0.15cm}AG & \vspace{0.15cm}H & \vspace{0.15cm}EG & \vspace{0.15cm}AG & \vspace{0.15cm}AG\\
        
        \includegraphics[scale=0.0875]{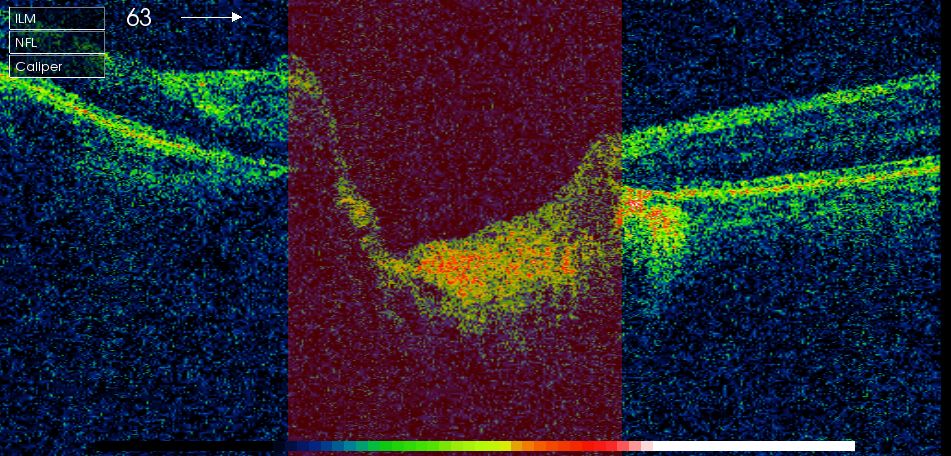} & \includegraphics[scale=0.04375]{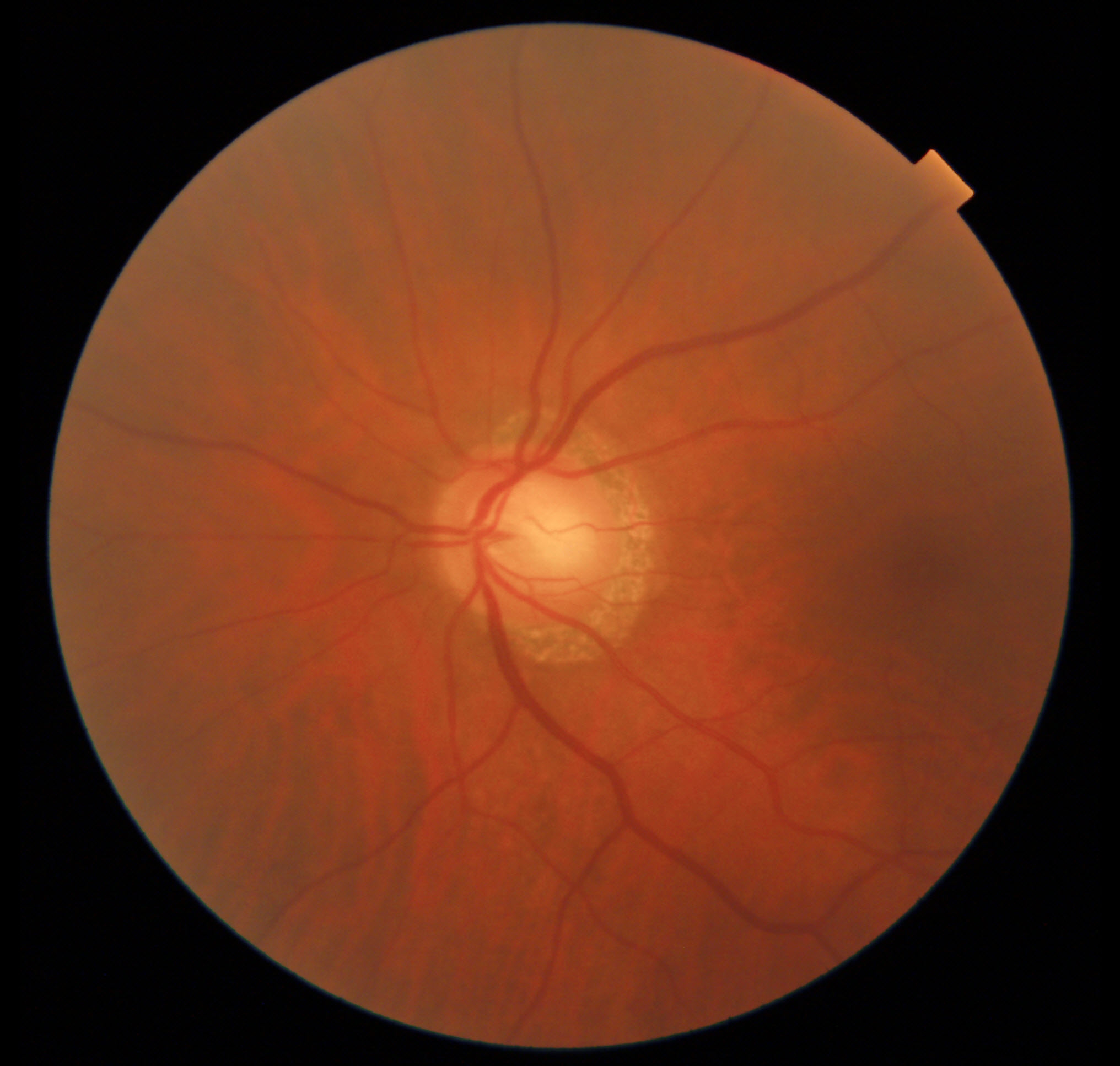} & AG & H & EG & AG & AG\\
        
        \includegraphics[scale=0.0875]{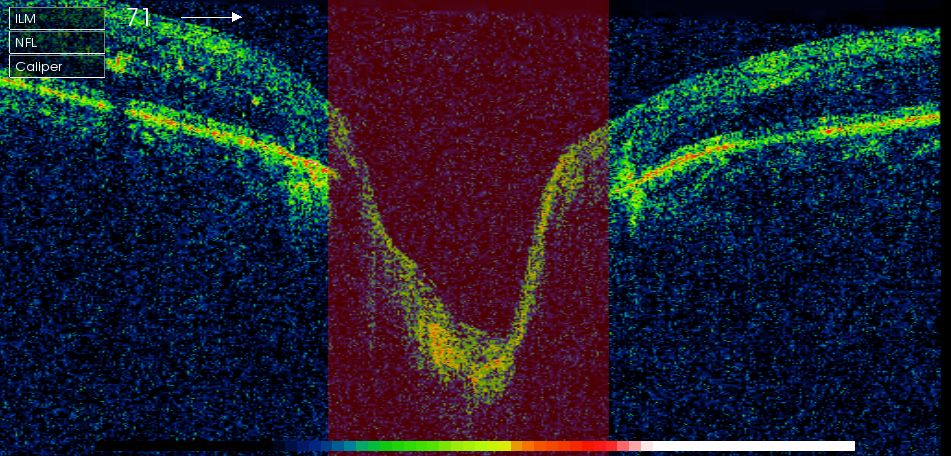} & \includegraphics[scale=0.04375]{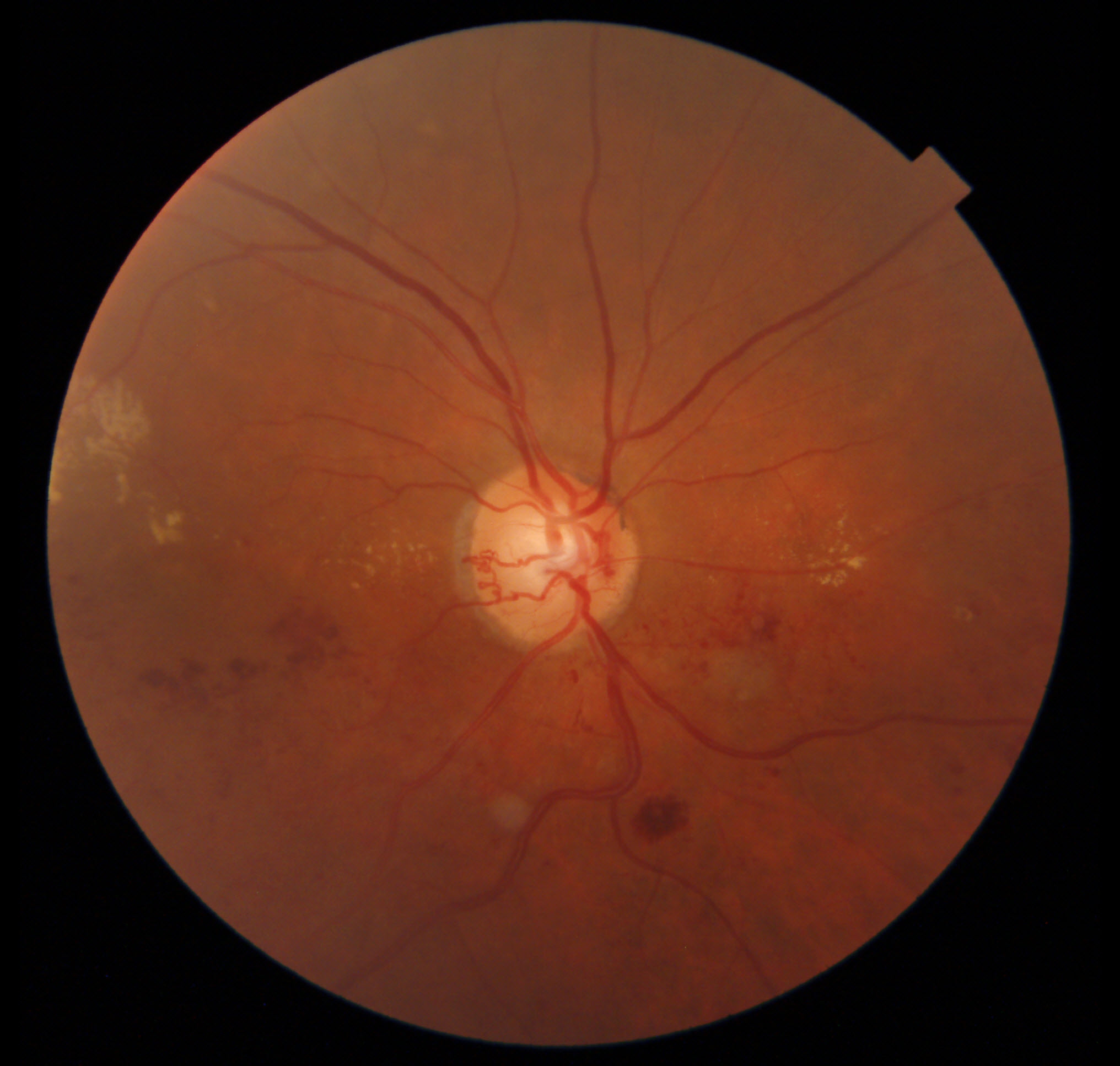} & EG & EG & EG & EG & EG\\
        
        \includegraphics[scale=0.0875]{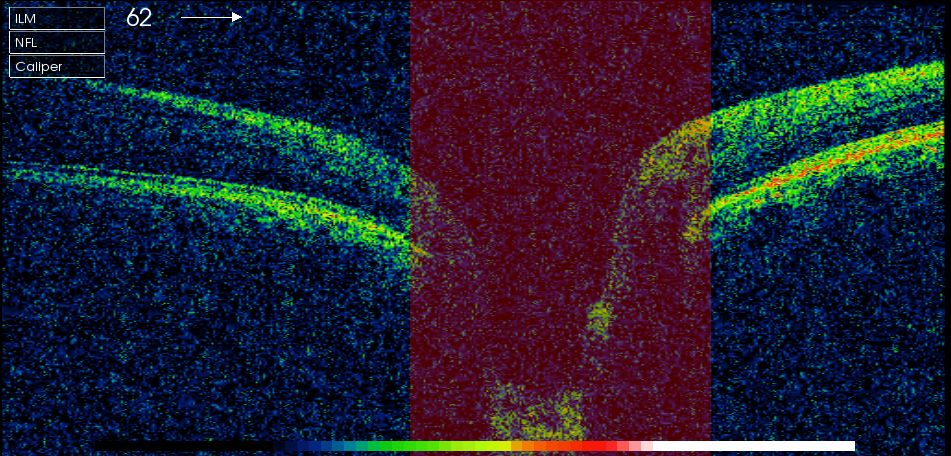} & \includegraphics[scale=0.04375]{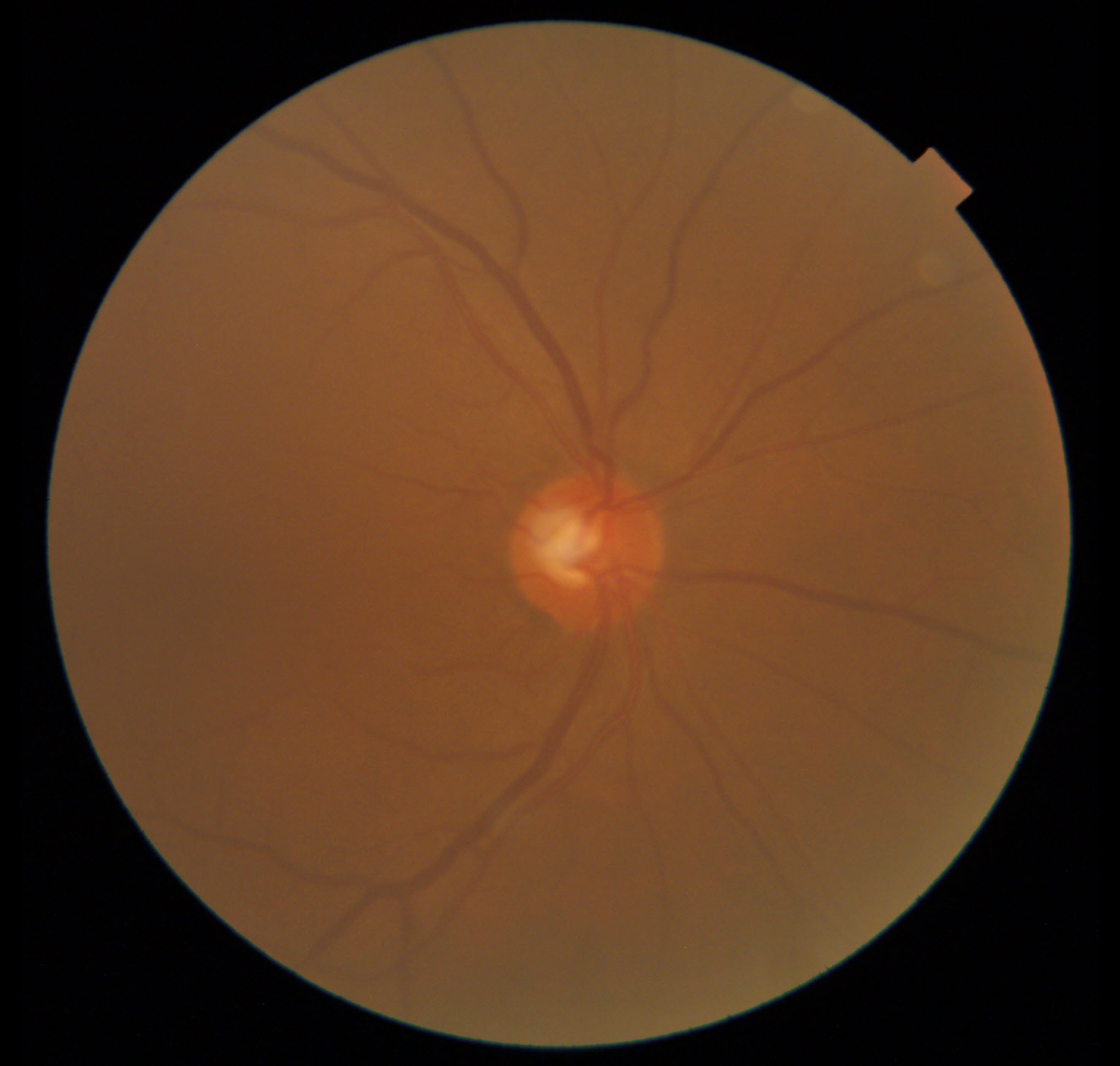} & EG & EG & EG & EG & EG\\
        
        \includegraphics[scale=0.0875]{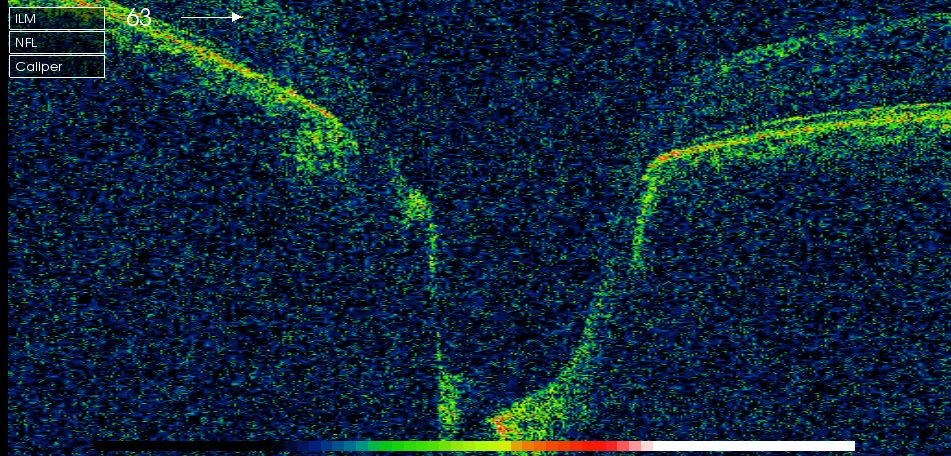} & \includegraphics[scale=0.04375]{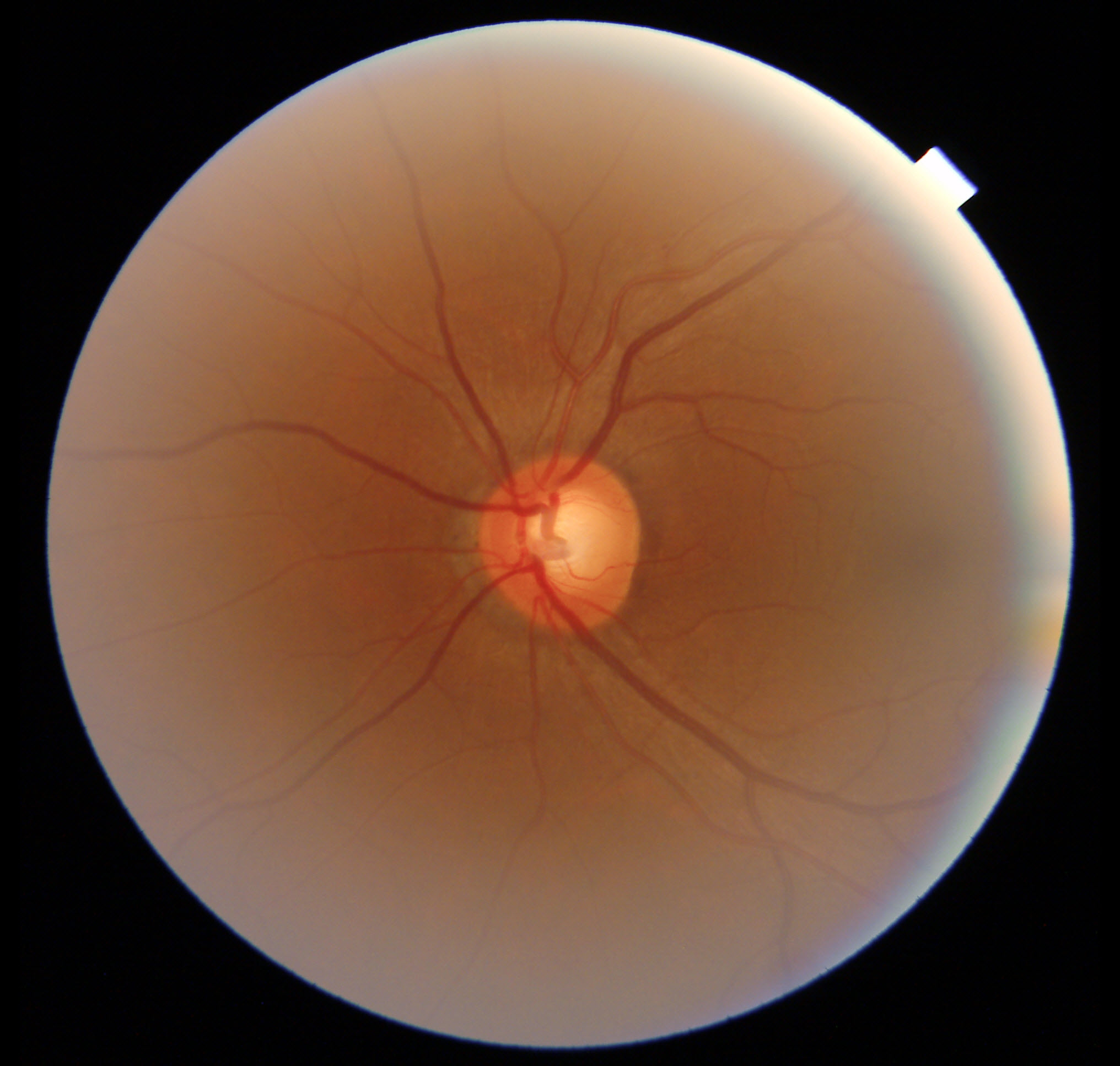} & EG & EG & H & EG & EG\\
        
        \includegraphics[scale=0.0875]{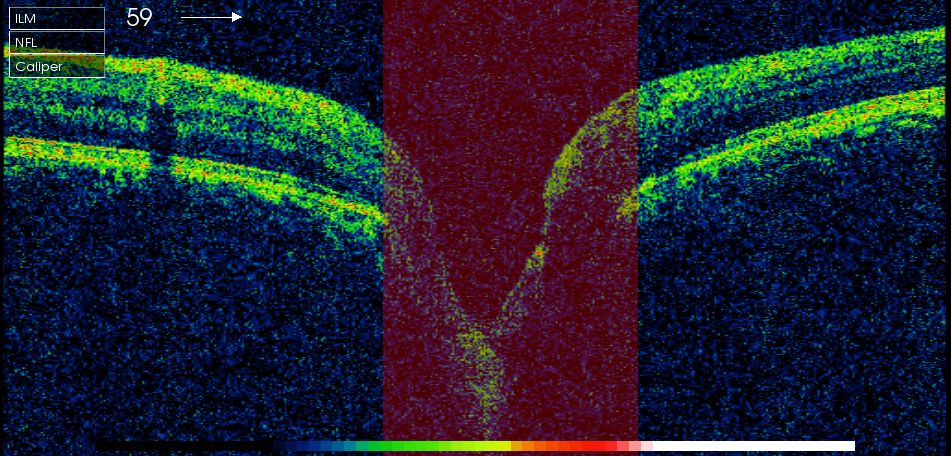} & \includegraphics[scale=0.04375]{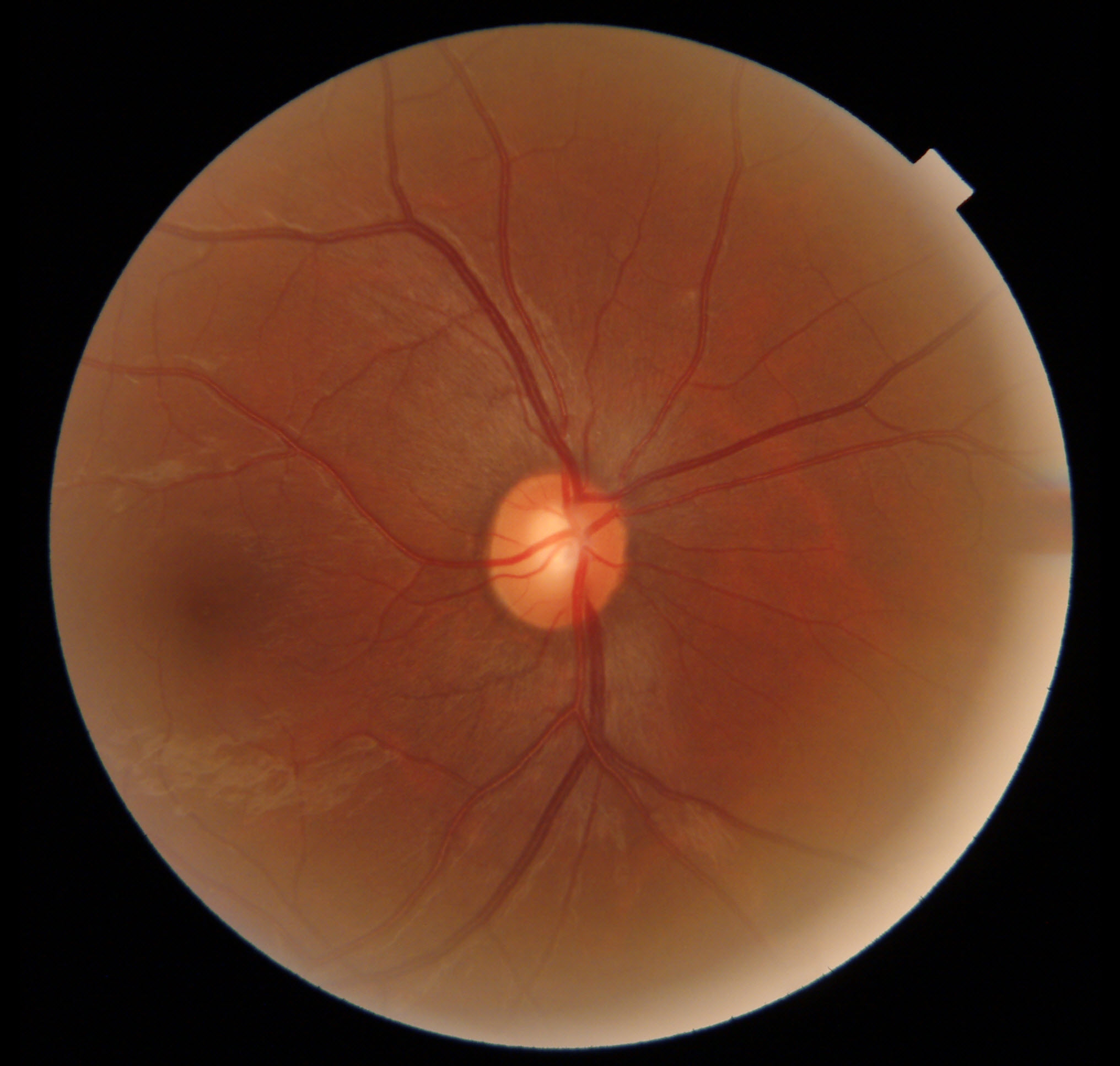} & H & H & H & H & H\\
        
        \includegraphics[scale=0.0875]{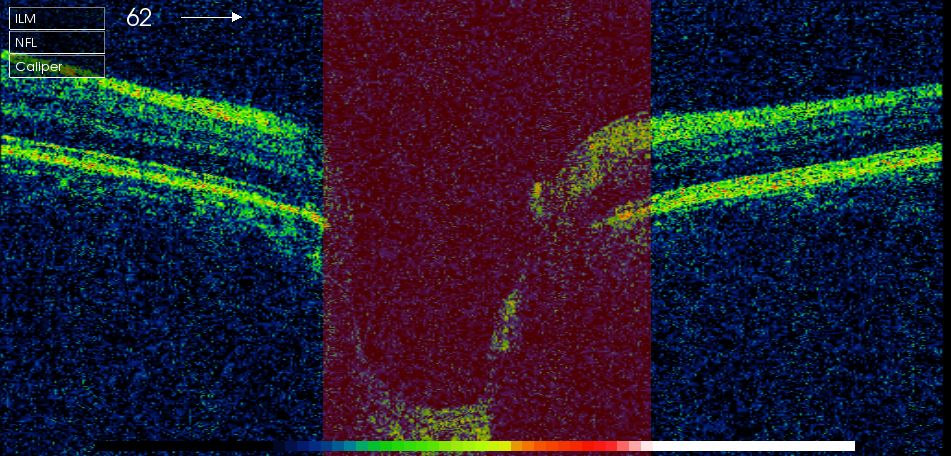} & \includegraphics[scale=0.04375]{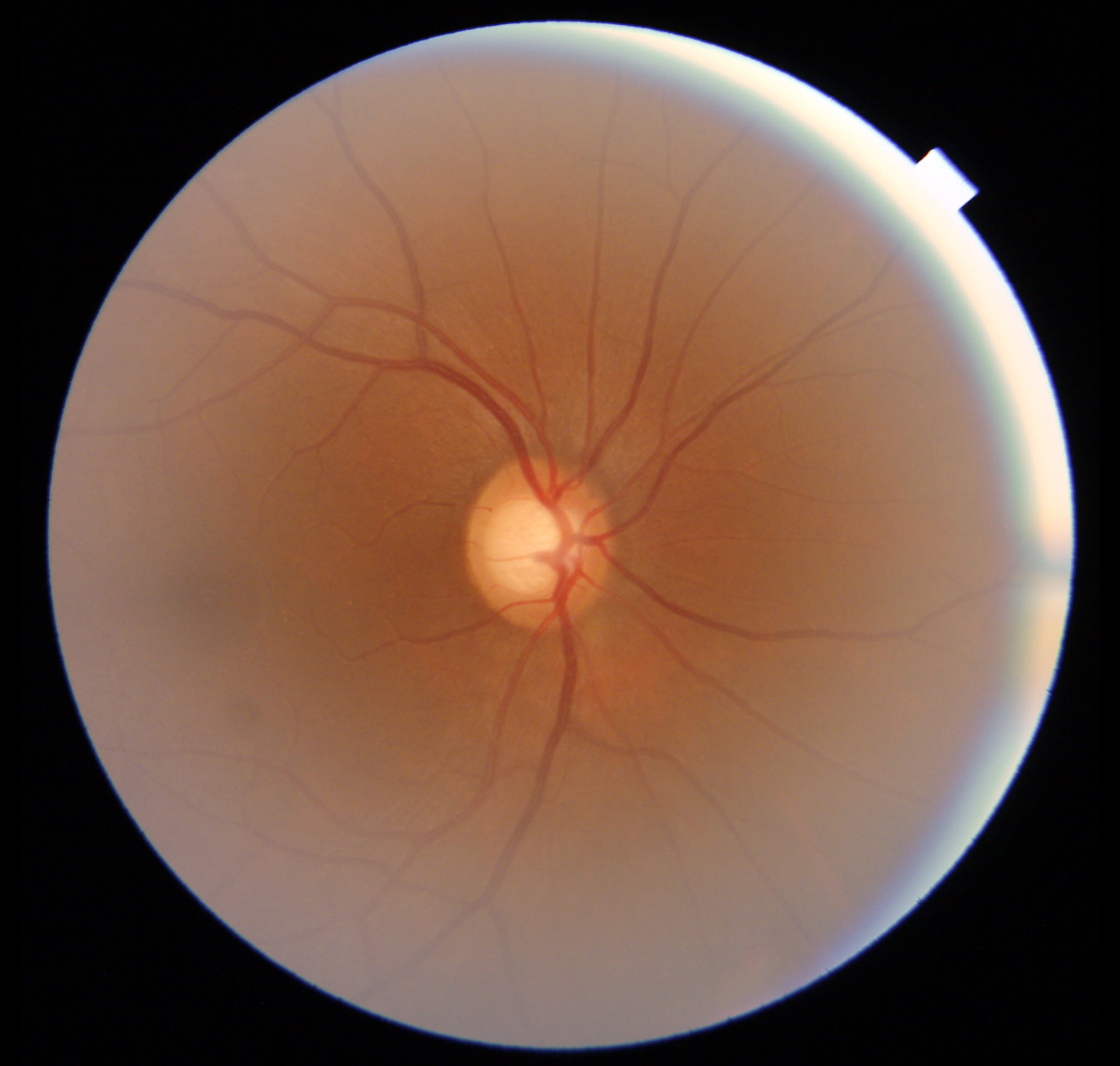} & AG & AG & AG & AG & AG\\
        
        \includegraphics[scale=0.0875]{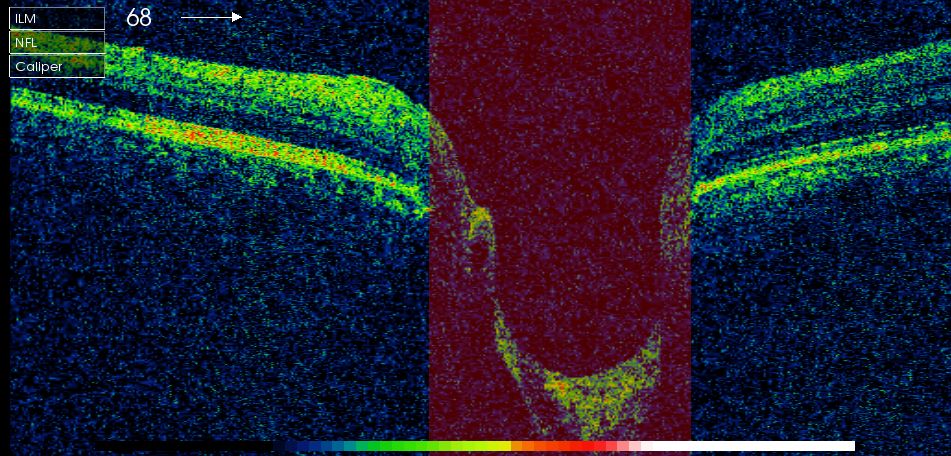} & \includegraphics[scale=0.04375]{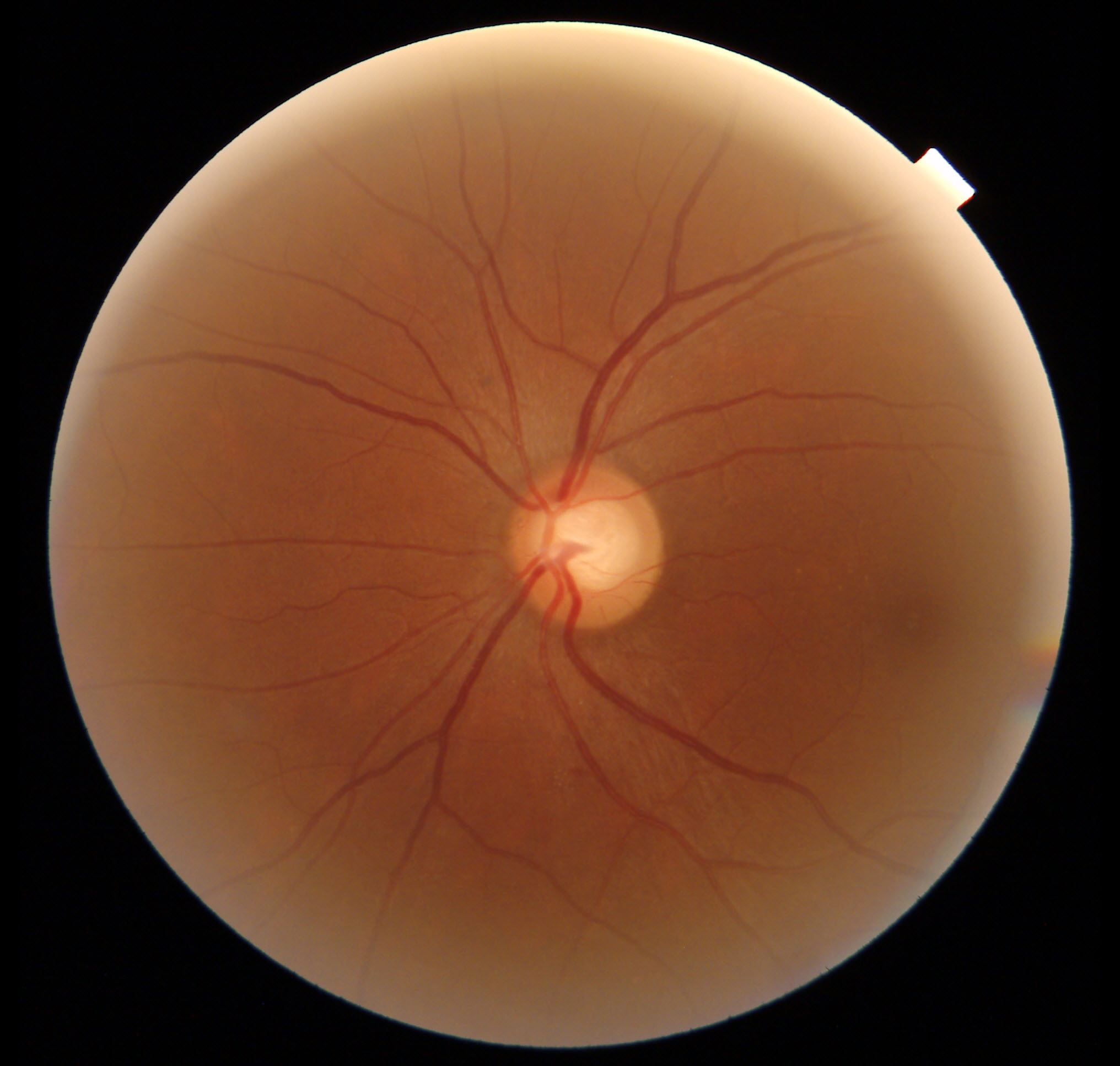} & EG & EG & EG & EG & AG\\
        
        \includegraphics[scale=0.0875]{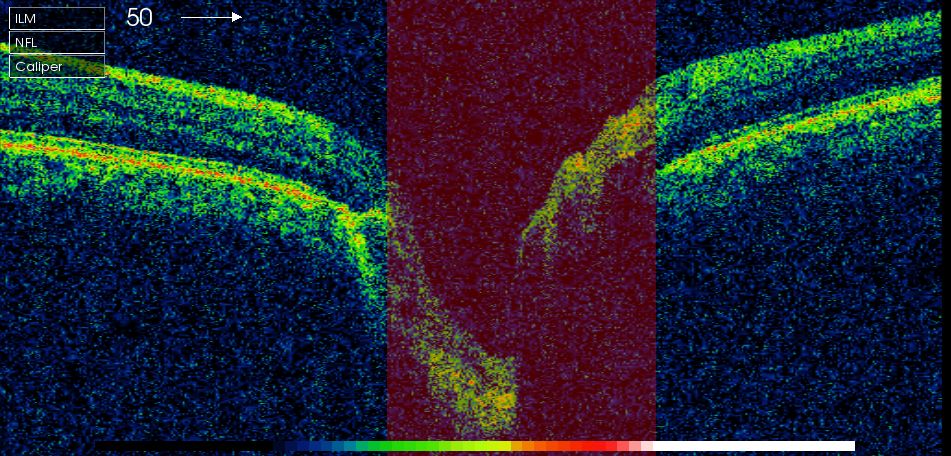} & \includegraphics[scale=0.04375]{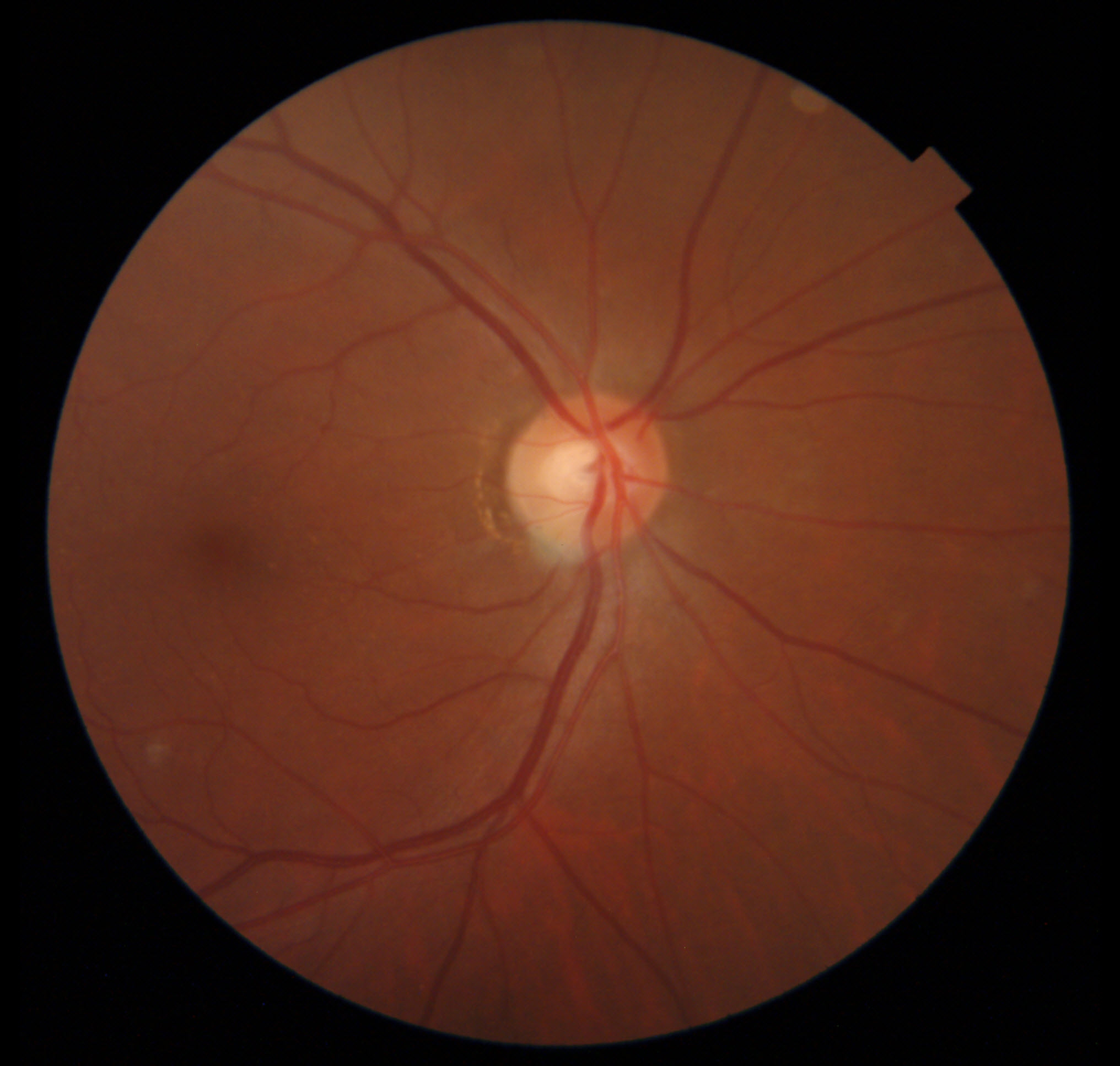} & H & H & EG & H & H\\
        
        \includegraphics[scale=0.0875]{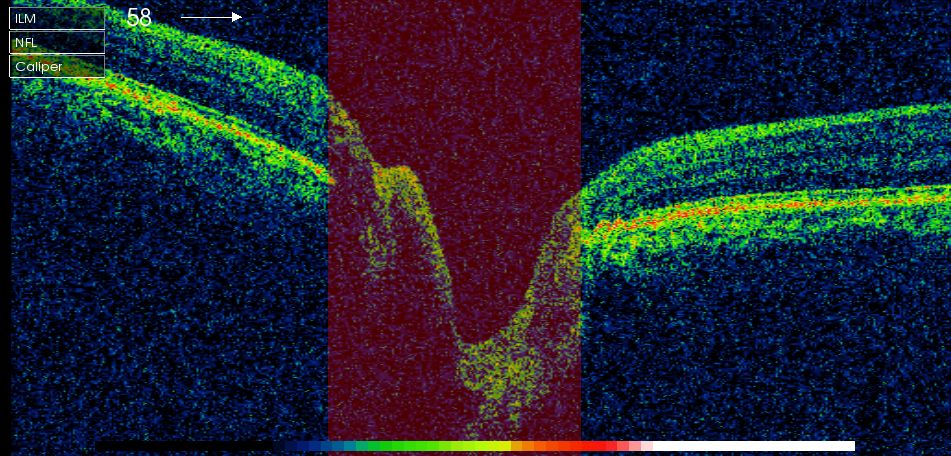} & \includegraphics[scale=0.04375]{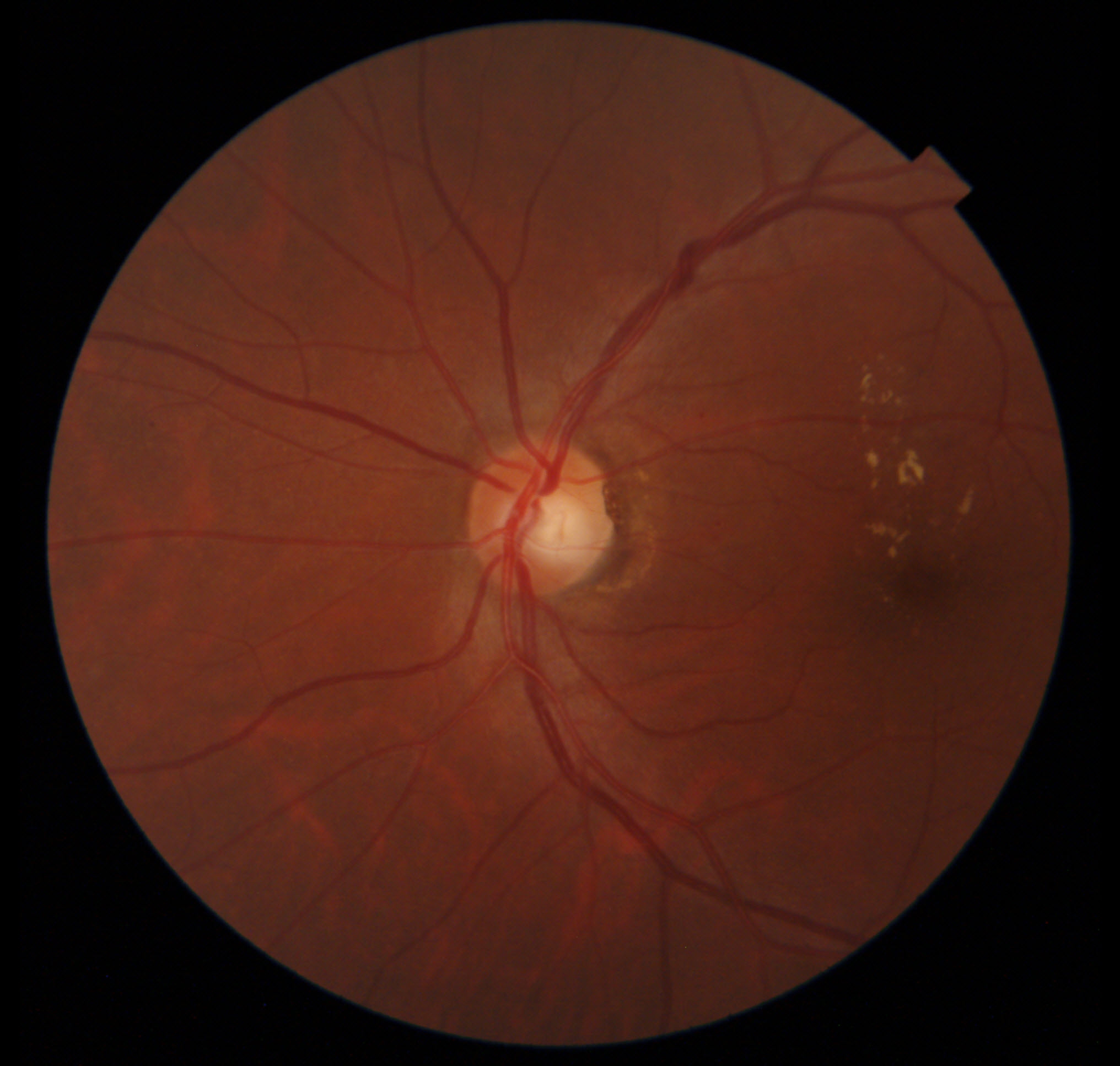} & H & H & H & H & H\\
        \bottomrule
    \end{tabular}
    \label{tab:tab7}
\end{table}

\begin{table}[htb]
    \centering
    \caption{Clinical validation of the proposed framework through the Pearson correlation coefficient ($r_c$) and its statistical significance ($p$-value).}
    \begin{tabular}{ccccc}
        \toprule
         Metric & Clinician-1 & Clinician-2 & Clinician-3 &Clinician-4  \\ \hline
         $r_c$ & 0.7260 &	0.8416 &	0.6863 &	0.9236 \\
         $p$-value & 1.04 $\times$ 10\textsuperscript{-23}  & 6.10 $\times$ 10\textsuperscript{-38} & 2.10 $\times$ 10\textsuperscript{-20} & 4.40 $\times$ 10\textsuperscript{-58}\\
         \bottomrule
    \end{tabular}
    \label{tab:tab10}
\end{table}

\section{Discussion and Conclusion} \label{sec:discussion}
\noindent In this work, we proposed a fully automated system for the classification and grading of glaucoma from ONH SD-OCT scans. Unlike other frameworks that rely on cup-to-disc ratios for screening glaucoma, the proposed system analyzes the pathological changes related to the degeneration of RGCs through RNFL, GC-IPL, and GCC thickness profiles. We propose a hybrid convolutional network (RAG-Net\textsubscript{v2})  for the extraction of these profiles, coupled with an SVM classifier for the classification and the grading of healthy and glaucomatous pathologies. The experiments evidenced the superiority of our framework in screening early and advanced glaucoma cases as compared to the state-of-the-art solutions relying on the cup-to-disc ratios as evidenced by the $F_1$ score of 0.9577. The preeminence of our system emanated from the newly proposed architecture variants in RAG-Net\textsubscript{v2}, integrating contextual-aware modules, built on residual atrous convolutions, along with the feature pyramid block. This proposed variants boosted the capacity of the RAG-Net\textsubscript{v2} for discriminating the retinal regions as reflected by the $\mu_{DC}$ score of 0.8697, outperforming popular deep segmentation models. Apart from this, RAG-Net\textsubscript{v2} significantly reduces the total number of trainable and non-trainable parameters by 91.04\%  as compared to the original RAG-Net architecture. This improvement also relates to the addition of contextual-aware modules replacing the standard convolutional blocks with the atrous convolutions, making the RAG-Net\textsubscript{v2} a lightweight architecture for screening and monitoring the glaucoma progression.
The introduction of contextual-aware modules in RAG-Net\textsubscript{v2}  addresses the limitations of the original RAG-Net architecture which, while outperforms state-of-the-art frameworks for lesions extraction,  showed restraints in differentiating similar textural patterns in retinal layers and boundaries as shown Figure \ref{fig:fig6}.  However, the implications of introducing contextual-aware modules need to be thoroughly tested for the lesions extraction from the macular pathologies, if we want to extend the modified RAG-Net architecture to the analysis of lesion-aware maculopathy. This investigation will be part of our next future work. 

%------------------------

\section*{Acknowledgment}
\noindent{This work is supported by a research fund from Khalifa University: Ref: CIRA-2019-047. We are also thankful to the four expert ophthalmologists for providing the clinical validations of the retinal scans within the AFIO dataset.} 
%%%%%%%%%%%%%%%%%%%%%%%%%%%%%%%%%%%%%%%%%%%%%%%%%%%%%%%%%%%%%%%%%%%%%%%%%%%%%%%%%%%%


\begin{thebibliography}{10}

\bibitem{weinreb2014JAMA}
R.~N. Weinreb, T.~Aung, and F.~A. Medeiros, ``The {P}athophysiology and
  {T}reatment of {G}laucoma: {A} {R}eview,''
\newblock JAMA, vol. 311, no. 18, pp. 1901-1911, 2014.

\bibitem{Burgoyne2005ONH}
C.~F. Burgoyne, J.~C. Downs, A.~J. Bellezza, J.-K.~F. Suh, and R.~T. Hart,
  ``{T}he optic nerve head as a biomechanical structure: A new paradigm for
  understanding the role of {IOP}-related stress and strain in the
  pathophysiology of glaucomatous optic nerve head damage,''
\newblock Progress in Retinal and Eye Research, vol 24, pp. 39-73, 2005. 

\bibitem{Raja2020DIB}
H.~Raja, M.~U. Akram, S.~G. Khawaja, M.~Arslan, A.~Ramzan, and N.~Nazir, ``Data
  on {O}{C}{T} and fundus images for the detection of glaucoma,''
\newblock Data in Brief, Volume 29, April 2020.

\bibitem{Hassan2015IST}
T.~Hassan, M.~U. Akram, B.~Hassan, A.~Nasim, and S.~A. Bazaz, ``{R}eview of
  {O}{C}{T} and fundus images for detection of {M}acular {E}dema,''
\newblock IEEE 12th International Conference on Imaging Systems and Techniques,
  2015.

\bibitem{majoor2019TVST}
J.~E.~A. Majoor, K.~A. Vermer, E.~R. Andrinopoulou, and H.~G. Lemij,
  ``{C}ontrast-to-{N}oise {R}atios for {A}ssessing the {D}etection of
  {P}rogression in the {V}arious {S}tages of {G}laucoma,''
\newblock Translational Vision Science \& Technology, vol. 8, no. 3, pp. 1-12,
  2019.

\bibitem{grewal2012Glaucoma}
D.~S. Grewal, M.~Sehi, J.~D. Paauw, and D.~S. Greenfield, ``{D}etection of
  {P}rogressive {R}etinal {N}erve {F}iber {L}ayer {T}hickness {L}oss {W}ith
  {O}ptical {C}oherence {T}omography {U}sing 4 {C}riteria for {F}unctional
  {P}rogression,''
\newblock Wolters Kluwer Journal of Glaucoma.

\bibitem{Gracitelli2015Ophthalmol}
M.~G. Wollstein, M.~J.~S. Schuman, M.~L.~L. Price, M.~A. Aydin, S.~P.~C. Stark,
  M.~E. Hertzmark, M.~E. Lai, M.~H. Ishikawa, M.~C. Mattox, P.~J.~G. Fujimoto,
  and P.~L.~A. Paunescu, ``{S}pectral-{D}omain {O}ptical {C}oherence
  {T}omography for {G}laucoma {D}iagnosis,''
\newblock Open Ophthalmol J, vol. 9, pp. 68-77, 2015.

\bibitem{leung2012Ophthalmology}
C.~K.~S. Leung, M.~Yu, R.~N. Weinreb, G.~Lai, G.~Xu, and D.~S.~C. Lam,
  ``Retinal nerve fiber layer imaging with spectral-domain optical coherence
  tomography: patterns of retinal nerve fiber layer progression,''
\newblock Ophthalmology, vol. 119, no. 9, pp. 1856-66, 2012.

\bibitem{ojima2007Ophthalmol}
T.~Ojima, T.~Tanabe, M.~Hangai, S.~Yu, S.~Morishita, and N.~Yoshimura,
  ``Measurement of {R}etinal {N}erve {F}iber {L}ayer {T}hickness and {M}acular
  {V}olume for {G}laucoma {D}etection {U}sing {O}ptical {C}oherence
  {T}omography,''
\newblock Jpn J Ophthalmol 2007, vol. 51, pp. 197-203, 2007.

\bibitem{Medeiros2012IOVS}
F.~A. Medeiros, L.~M. Zangwill, C.~Bowd, K.~Mansouri, and R.~N. Weinreb, ``The
  {S}tructure and {F}unction {R}elationship in {G}laucoma: {I}mplications for
  {D}etection of {P}rogression and {M}easurement of {R}ates of {C}hange,''
\newblock Investigative Ophthalmology \& Visual Science, vol. 53, no. 11, pp.
  6939-6946, 2012.

\bibitem{El-Naby2014Egyptian}
A.~E.~A. El-Naby, H.~Y. Abouelkheir, H.~T. Al-Sharkawy, and T.~H. Mokbel,
  ``Correlation of retinal nerve fiber layer thickness and perimetric changes
  in primary open-angle glaucoma,''
\newblock Journal of the Egyptian Ophthalmological Society, Vol. 111, Issue 1,
  2018.

\bibitem{Khalil2020Wiley}
T.~Shehryar, M.~U. Akram, S.~Khalid, S.~Nasreen, A.~Tariq, A.~Perwaiz, and
  A.~Shaukat, ``Improved automated detection of glaucoma by correlating fundus and {S}{D}‐{O}{C}{T} image analysis,''
\newblock Int J Imaging Syst Technol., 1– 20, 2020.

\bibitem{Sun2018Localizing}
X.~Sun, Y.~Xu, M.~Tan, H.~Fu, W.~Zhao, T.~You, and J.~Liu, ``{L}ocalizing
  {O}ptic {D}isc and {C}up for {G}laucoma {S}creening via {D}eep {O}bject
  {D}etection {N}etworks,''
\newblock Computational Pathology and Ophthalmic Medical Image Analysis,
  Springer, Cham, pp. 236-244, 2018.

\bibitem{Chen2015Glaucoma}
X.~Chen, Y.~Xu, D.~W.~K. Wong, T.~Y. Wong, and J.~Liu, ``Glaucoma detection
  based on deep convolutional neural network,''
\newblock 37th Annual International Conference of the IEEE Engineering in
  Medicine and Biology Society (EMBC), 2015.

\bibitem{Cheng2013Superpixel}
J.~Cheng, J.~Liu, Y.~Xu, F.~Yin, D.~W.~K. Wong, N.-M. Tan, D.~Tao, C.-Y. Cheng,
  T.~Aung, and T.~Y. Wong, ``{S}uperpixel {C}lassification {B}ased {O}ptic
  {D}isc and {O}ptic {C}up {S}egmentation for {G}laucoma {S}creening,''
\newblock IEEE Transactions on Medical Imaging, vol 32, issue 6, pp. 1019-1032,
  2013.

\bibitem{Fu2018DiscAware}
H.~Fu, J.~Cheng, Y.~Xu, C.~Zhang, D.~W.~K. Wong, J.~Liu, and X.~Cao,
  ``{D}isc-{A}ware {E}nsemble {N}etwork for {G}laucoma {S}creening {F}rom
  {F}undus {I}mage,''
\newblock IEEE Transactions on Medical Imaging, vol 37, issue 11, pp.
  2493-2501, 2018.

\bibitem{Khalil2018Access}
T.~Khalil, M.~U. Akram, H.~Raja, A.~Jameel, and I.~Basit, ``{I}mproved
  automated detection of glaucoma from fundus image using hybrid structural and
  textural features,''
\newblock IEEE Access, Vol. 6, pp. 4560-4576, 2018.

\bibitem{Almobarak2014IOVS}
F.~A. Almobarak, N.~O'Leary, A.~S.~C. Reis, G.~P. Sharpe, D.~M. Hutchison,
  M.~T. Nicolela, and B.~C. Chauhan, ``Automated segmentation of optic nerve
  head structures with optical coherence tomography,''
\newblock Invest Ophthalmol Vis Sci., 26, 55(2), 1161-8, 2014.

\bibitem{Kromer2017Ophthalmology}
R.~Kromer, S.~Rahman, F.~Filev, and M.~Klemm, ``{A}n {A}pproach for {A}utomated
  {S}egmentation of {R}etinal {L}ayers {I}n {P}eripapillary {S}pectralis
  {SDOCT} {I}mages {U}sing {C}urve {R}egularisation,''
\newblock Insights in Ophthalmology, vol. 1, no. 10, pp. 1-6, 2017.

\bibitem{Duan2018Access}
W.~Duan, Y.~Zheng, Y.~Ding, S.~Hou, Y.~Tang, Y.~Xu, M.~Qin, J.~Wu, and D.~Shen,
  ``{A} {G}enerative {M}odel for {OCT} {R}etinal {L}ayer {S}egmentation by
  {G}roupwise {C}urve {A}lignment,''
\newblock IEEE Access, vol. 6, pp. 25130-25141, 2018.

\bibitem{Niua2014CBM}
S.~Niu, Q.~Chena, L.~D. Sisternesb, D.~L. Rubinb, W.~Zhangc, and Q.~Liu,
  ``Automated retinal layers segmentation in {S}{D}-{O}{C}{T} images using
  dual-gradient and spatial correlation smoothness constraint,''
\newblock Computers in Biology and Medicine, vol. 54, pp. 116–128, 2014.

\bibitem{Kafieh2015Hindawi}
R.~Kafieh, H.~Rabbani, F.~Hajizadeh, M.~D. Abramoff, and M.~Sonka,
  ``{T}hickness {M}apping of {E}leven {R}etinal {L}ayers {S}egmented {U}sing
  the {D}iffusion {M}aps {M}ethod in {N}ormal {E}yes,''
\newblock Hindawi Journal of Ophthalmology, vol. 2015, p. 14. Article ID
  259123, March 2015.

\bibitem{Bagci2007LSSAW}
A.~M. Bagci, R.~Ansari, and M.~Shahid, ``{A} {M}ethod for {D}etection of
  {R}etinal {L}ayers by {O}ptical {C}oherence {T}omography {I}mage
  {S}egmentation,''
\newblock IEEE/NIH Life Science Systems and Applications Workshop, 2007.

\bibitem{Abdellatif2019Hindawi}
M.~K. Abdellatif, Y.~A.~M. Elzankalony, A.~A.~A. Ebeid, and W.~M. Ebeid,
  ``{O}uter {R}etinal {L}ayers’ {T}hickness {C}hanges in relation to {A}ge
  and {C}horoidal {T}hickness in {N}ormal {E}yes,''
\newblock Hindawi Journal of Ophthalmology, vol. 2019, pp. 8, Article ID
  1698967, July 2019.

\bibitem{Hassan2016AO}
T.~Hassan, M.~U. Akram, B.~Hassan, A.~M. Syed, and S.~A. Bazaz, ``Automated
  segmentation of subretinal layers for the detection of macular edema,''
\newblock Applied Optics Vol. 55, Issue 3, pp. 454-461, 2016.

\bibitem{Hassan2019CBM}
T.~Hassan, M.~U. Akram, M.~F. Masood, and U.~Yasin, ``Deep structure tensor
  graph search framework for automated extraction and characterization of
  retinal layers and fluid pathology in retinal {S}{D}-{O}{C}{T} scans,''
\newblock Computers in Biology and Medicine, Volume 105, Pages 112-124,
  February 2019.

\bibitem{Hassan2018Access}
T.~Hassan, M.~U. Akram, A.~Shaukat, S.~G. Khawaja, and B.~Hassan, ``{S}tructure
  {T}ensor {G}raph {S}earches {B}ased {F}ully {A}utomated {G}rading and
  {P}rofiling of {M}aculopathy {F}rom {R}etinal {O}{C}{T} {I}mages,''
\newblock IEEE Access, vol. 6, pp. 44644 - 44658, 2018.

\bibitem{Chiu2015BOE}
S.~J. Chiu, M.~J. Allingham, P.~S. Mettu, S.~W. Cousins, J.~A. Izatt, and
  S.~Farsiu, ``Kernel regression based segmentation of optical coherence
  tomography images with diabetic macular edema,''
\newblock Biomedical Optics Express, 6(4), pp. 1172–1194, 2015.

\bibitem{Ometto2019Trans}
G.~Ometto, I.~Moghul, G.~Montesano, A.~Hunter, N.~Pontikos, P.~R. Jones, P.~A.
  Keane, X.~Liu, A.~K. Denniston, and D.~P. Crabb, ``{R}e{L}ayer: a {F}ree,
  {O}nline {T}ool for {E}xtracting {R}etinal {T}hickness {F}rom
  {C}ross-{P}latform {O}{C}{T} {I}mages,''
\newblock Trans Vis Sci Tech, vol. 8, no. 3, p. Article 25, 2019.

\bibitem{Gao2015PLOSONE}
E.~Gao, B.~Chen, J.~Yang, F.~Shi, W.~Zhu, D.~Xiang, H.~Chen, M.~Zhang, and
  X.~Chen., ``{C}omparison of {R}etinal {T}hickness {M}easurements between the
  {T}opcon {A}lgorithm and a {G}raph-{B}ased {A}lgorithmin {N}ormal and
  {G}laucoma {E}yes,''
\newblock PLOS ONE, vol. 10, no. 6, p. e0128925., 2015.

\bibitem{Mayer2010BOE}
M.~A. Mayer, J.~Hornegger, and T.-R.~P. Mardin, C.~Y., ``{R}etinal {N}erve
  {F}iber {L}ayer {S}egmentation on {F}{D}-{O}{C}{T} {S}cans of {N}ormal
  {S}ubjects and {G}laucoma {P}atients,''
\newblock Biomedical Optics Express, vol. 1, no. 5, pp. 1358-1381, 2010.

\bibitem{Khalil2017IET}
T.~Khalil, M.~U. Akram, S.~Khalid, and A.~Jameel, ``{I}mproved automated
  detection of glaucoma from fundus image using hybrid structural and textural
  features,''
\newblock IET Image Processing, Volume: 11 , Issue: 9, 2017.

\bibitem{Mariottoni2020ScientificReports}
E.~B. Mariottoni, A.~A. Jammal, C.~N. Urata, S.~I. Berchuck, A.~C. Thompson,
  T.~Estrela, and F.~A. Medeiros, ``{Q}uantification of {R}etinal {N}erve
  {F}ibre {L}ayer {T}hickness on {O}ptical {C}oherence {T}omography with a
  {D}eep {L}earning {S}egmentation-{F}ree {A}pproach,''
\newblock Scientific Reports, vol. 10, p. Article 402, 2020.

\bibitem{Zang2019BOE}
P.~Zang, J.~Wang, T.~T. Hormel, L.~Liu, D.~Huang, and Y.~Jia, ``Automated
  segmentation of peripapillary retinal boundaries in oct combining a
  convolutional neural network and a multi-weights graph search,''
\newblock Biomedical Optics Express, Vol. 10, Issue 8, pp. 4340-4352, 2019.

\bibitem{Maetschke2019PLOSONE}
S.~Maetschke, B.~Antony, H.~Ishikawa, G.~Wollstein, J.~Schuman, and R.~Garnavi,
  ``A feature agnostic approach for glaucoma detection in oct volumes,''
\newblock PLOS ONE, 14(7): e0219126, 2019.

\bibitem{Sripad2018BOE}
S.~K. Devalla, P.~K. Renukanand, B.~K. Sreedhar, G.~Subramanian, L.~Zhang,
  S.~Perera, J.-M. Mari, K.~S. Chin, T.~A. Tun, N.~G. Strouthidis, T.~Aung,
  A.~H. Thiéry, and M.~J.~A. Girard, ``{D}{R}{U}{N}{E}{T}: a dilated-residual
  {U}-{N}et deep learning network to segment optic nerve head tissues in
  optical coherence tomography images,''
\newblock Biomedical Optics Express, Vol. 9, Issue 7, pp. 3244-3265, 2018.

\bibitem{Wang2019BOE}
J.~Wang, Z.~Wang, F.~Li, G.~Qu, Y.~Qiao, H.~Lv, and X.~Zhang, ``Joint retina
  segmentation and classification for early glaucoma diagnosis,''
\newblock Biomedical Optics Express, Vol. 10, Issue 5, pp. 2639-2656, 2019.

\bibitem{st2}
J.~Bigun, G.~Granlund, and J.~Wiklund, ``{M}ultidimensional {O}rientation
  {E}stimation with {A}pplications to {T}exture {A}nalysis and {O}ptical
  {F}low,''
\newblock IEEE Transactions on Pattern Analysis and Machine Intelligence, 1991.

\bibitem{Hassan2020JBHI}
T.~Hassan, M.~U. Akram, N.~Werghi, and N.~Nazir, ``{R}{A}{G}-{F}{W}: {A} hybrid
  convolutional framework for the automated extraction of retinal lesions and
  lesion-influenced grading of human retinal pathology,''
\newblock IEEE Journal of Biomedical and Health Informatics,
  10.1109/JBHI.2020.2982914, March 2020.

\bibitem{Hassan2020}
T.~Hassan, M.~U. Akram, and N.~Werghi, ``{E}valuation of {D}eep {S}egmentation
  {M}odels for the {E}xtraction of {R}etinal {L}esions from {M}ulti-modal
  {R}etinal {I}mages,''
\newblock arXiv:2006.02662, June 2020.

\bibitem{Wang2018KDD}
Z.~Wang and S.~Ji, ``{S}moothed {D}ilated {C}onvolutions for {I}mproved {D}ense
  {P}rediction,''
\newblock 24TH ACM Sigkdd Conference On Knowledge Discovery And Data Mining,
  2018.

\bibitem{Wang2018WACV}
P.~Wang, P.~Chen, Y.~Yuan, D.~Liu, Z.~Huang, X.~Hou, and G.~Cottrell,
  ``Understanding convolution for semantic segmentation,''
\newblock IEEE Winter Conference on Applications of Computer Vision (WACV),
  2018.

\bibitem{Zeiler2012ADADELTA}
M.~D. Zeiler, ``{A}{D}{A}{D}{E}{L}{T}{A}: {A}n {A}daptive {L}earning {R}ate
  {M}ethod,''
\newblock arXiv:1212.5701, 2012.

\bibitem{PSPNet}
H.~Zhao, J.~Shi, X.~Qi, X.~Wang, and J.~Jia, ``{P}yramid {S}cene {P}arsing
  {N}etwork,''
\newblock IEEE International Conference on Computer Vision and Pattern
  Recognition, 2017.

\bibitem{SegNet}
V.~Badrinarayanan, A.~Kendall, and R.~Cipolla, ``Seg{N}et: {A} {D}eep
  {C}onvolutional {E}ncoder-{D}ecoder {A}rchitecture for {I}mage
  {S}egmentation,''
\newblock IEEE Transactions on Pattern Analysis and Machine Intelligence,
  December 2017.

\bibitem{unet}
O.~Ronneberger, P.~Fischer, and T.~Brox, ``U-{N}et: {C}onvolutional {N}etworks
  for {B}iomedical {I}mage {S}egmentation,''
\newblock Medical Image Computing and Computer Assisted Intervention (MICCAI),
  2015.

\bibitem{fcn}
J.~Long, E.~Shelhamer, and T.~Darrell, ``{F}ully {C}onvolutional {N}etworks for
  {S}emantic {S}egmentation,'' in {\em IEEE International Conference on
  Computer Vision and Pattern Recognition (CVPR),}
\newblock 2015.

\bibitem{Illinois}
``{E}ye {E}xam and tests for {G}laucoma diagnosis,''
\newblock The Eye Digest. The University of Illinois Eye and Ear Infirmary.
  Archived from the original on 8 July 2012.

\end{thebibliography}


\small
\end{document}